\documentclass[lettersize, journal]{IEEEtran}
\PassOptionsToPackage{table}{xcolor}
\ifCLASSOPTIONcompsoc
  \usepackage[nocompress]{cite}
\else
  \usepackage{cite}
\fi

\usepackage{graphicx} 
\usepackage{url}
\usepackage{booktabs}
\usepackage{multirow}
\usepackage{tcolorbox}
\usepackage{subcaption}
\usepackage{algorithm,algorithmic}
\usepackage{amsmath}
\usepackage{soul}
\usepackage{hyperref}

\usepackage[T1]{fontenc}

\usepackage{xcolor}

\newif\ifshowrevisions
\showrevisionsfalse       

\newcommand{\revision}[1]{%
  \ifshowrevisions
    {\color{blue}#1}%
  \else
    #1%
  \fi
}

\newcommand{\revisioncolor}{\ifshowrevisions\color{blue}\arrayrulecolor{black}\fi}

\begin{document}

\title{Does My README File Need To Be Updated? Exploring LLM-Based README Maintenance}

\author{Haoyu~Gao,
        Hong~Yi~Lin,
        Christoph~Treude,
        Gregory~Gay,
       Mansooreh~Zahedi
\thanks{Haoyu~Gao, Hong~Yi~Lin and Mansooreh~Zahedi are with the School of Computing and Information Systems, the University of Melbourne, Victoria, 3053
Australia (e-mail: haoyug1@student.unimelb.edu.au).}
\thanks{Gregory~Gay is with the Department of Computer Science \& Engineering, Chalmers University of Technology and University of Gothenburg, Gothenburg, 417 56 Sweden}
\thanks{Christoph~Treude is with the School of Computing and Information Systems, Singapore Management University,  178903 Singapore}}

\maketitle

\begin{abstract}

The README file serves as a critical information source for project overviews and developer onboarding in Open Source Software (OSS).  Yet, documentation issues persist. In particular, ``outdated'' documentation is perceived by developers as one of the most frequent and severe challenges with gaining project understanding. While previous studies aimed to mitigate this problem, they typically rely on highly engineered solutions for specific code components or employ generative methods ineffective for incremental maintenance. \revision{In this study, we formulate \emph{surgical documentation update recommendation} as a task, and present a Large Language Model (LLM)-driven framework as a proof of concept for it within a human-in-the-loop workflow.} Specifically, given a pull request (PR), \revision{the framework} determines whether an update is necessary, and if so, it identifies the locations where updates should be applied and provides a justification based on the triggering events.

\revision{Our evaluation on 25,511 PRs across 714 popular repositories establishes the feasibility of the task. Its best configuration recovers half of the PRs that were historically accompanied by a README update, and reaches a user-facing accuracy of 28\% under the observed prevalence of such updates.} Furthermore, we performed a qualitative failure case analysis to provide deeper insights and directions for improvement. We also conducted a retrospective study on 20 sampled repositories, complemented by a case study with a developer of a large OSS project. \revision{Manual annotation confirmed that 21.5\% of the temporally matched recommendations identified updates developers had overlooked, or 6.1\% under the most conservative reading, so the user-facing accuracy we report is a lower bound on what a deployment would see.} Finally, we provide concrete implications for practitioners and researchers, highlighting the need to further explore effective interaction patterns to incorporate documentation update tools into the OSS development workflow.

\end{abstract}

\section{Introduction}

Software documentation serves as a critical information source for diverse stakeholders in the software ecosystem~\cite{parnas2009document}. In particular, README file acts as the primary entry point to a project, existing in nearly all Open Source Software (OSS). It provides an introductory project overview~\cite{prana2019categorizing}, encompassing essential instructions, primarily ``What'' the software is and ``How'' to use it, alongside other categories such as ``Contribution'' guidelines, enabling users to rapidly grasp high-level knowledge of the repository. Consequently, README plays a pivotal role in forming an end user's first impression of the project~\cite{koskela2018open} and facilitating the onboarding of newcomers~\cite{steinmacher2016overcoming}.

However, software documentation suffers from numerous quality issues, including concerns regarding ``up-to-dateness'', ``correctness'', and ``presentation''~\cite{aghajani2019software}, all of which compromise its usability. Specifically, outdated documentation is perceived by practitioners as one of the most severe challenges, yet it remains an area with comparatively few effective solutions~\cite{aghajani2020software}. In response to the rapidly evolving software ecosystem, recent work has categorised the update behaviours and triggering events within README files. A taxonomy provides actionable guidance for identifying update triggers and planning necessary documentation revisions~\cite{gao2025adapting}. Meanwhile, automated approaches targeting documentation updates have been actively investigated. Previous research on detecting and updating outdated documentation primarily focused on leveraging specifically-engineered traceability links, such as mapping code components in the codebase to code-like terms in documents, to infer outdated elements~\cite{dagenais2014using, tan2024detecting}. However, these rigid approaches do not suit high-level documents such as a README, as the information presented is usually abstracted without a clear ``exact match'' to the codebase components. For instance, switching internal components such as authentication protocols will not be captured by an exact match of code components, and is consequently ignored by the aforementioned paradigm.

Additionally, with the advent of LLMs, the research focus has shifted toward automatically generating repository-level documentation, such as README files, from scratch~\cite{11186105, koreeda2023larch}. However, such generative approaches align poorly with the task of maintaining documentation, as they are often destructive---disregarding the existing document's content and style. Regenerating a full README file to reflect minor updates risks style drift and the incorporation of hallucinated content, mirroring the over-correction issue in grammatical error correction where correct segments are unnecessarily modified~\cite{li2025harnessing}. Furthermore, the entire codebase context incurs high computational costs and degrades reliability due to the ``Lost in the Middle'' phenomenon~\cite{liu2024lost}. Finally, whole-file regeneration imposes a significantly higher review burden on developers than reviewing targeted smaller edits~\cite{sadowski2018modern}.
\revision{Beyond these costs, whole-file regeneration is untargeted, in that it leaves the maintainer without any indication of which part of the document a given code change has invalidated, which is the question a maintainer reviewing a PR actually faces.}


\revision{Consequently, we formulate documentation maintenance in pull-based development as a distinct task, which we refer to as \emph{surgical documentation update recommendation}. Given a PR and a documentation artefact, the task is to decide whether the change invalidates any part of that artefact, and, if so, to return a ranked list of the specific sections that require revision together with a justification that links each one to the triggering change. The existing document is an input that is not modified, and the output is a located and explained recommendation rather than a replacement text. Section~\ref{sec:motivating_example} states the task formally, and our framework instantiates it by identifying fine-grained locations within the README that require modification and generating a justification based on the triggering event. Formulating the task this way retains the developer as the decision maker, since a located and justified recommendation can be weighed against project knowledge that an automated tool lacks.}

We utilised the dataset made available by Lin et al.~\cite{lin2025leveraging} to construct a ground truth of PRs with README updates and a control set of PRs \textit{without} README updates. Ground truth reflects observed developer behaviour, not necessarily optimal documentation quality. We conducted comprehensive end-to-end and component-wise evaluations at both the PR and paragraph levels, supplemented by manual verification of the correctness of intermediate reasoning and justifications. Under real-world distribution assumptions, our best approach achieves a user-facing accuracy of 28\%. While this number may appear modest in isolation, documentation updates are relatively rare events (less than 1\%) in PRs, meaning that developers would otherwise need to manually inspect hundreds of PRs to identify one that requires a README change. In practice, our results indicate that developers need to review fewer than four flagged recommendations to identify one valid update, substantially reducing the search effort compared to manual monitoring across the entire PR stream. 


Finally, we assessed real-world utility through a retrospective study on 20 repositories and an interview with a core developer of \texttt{jabref}, a popular open source project with nearly 900 contributors, 50 releases and more than 10,000 PRs. Manual analysis confirmed that 21.5\% of recommendations for PRs lacking immediate README updates were actually valid suggestions overlooked during development, which were validated as they were later updated in future development activities. \revision{A maintainer of one of these projects also judged recommendations to be useful in cases where no subsequent update was recorded, providing additional anecdotal evidence of the tool's capacity to mitigate documentation debt.} Because our tool successfully identifies valid updates that were missed in the historical ground truth, the actual user-facing accuracy is likely higher than our conservatively reported figures. Finally, we discuss the implications, highlighting the need for future research into interaction patterns for integrating documentation maintenance tools into the OSS development workflow.

This paper makes four contributions.
\begin{itemize}
  \item \revision{A formulation of surgical documentation update recommendation as a task, with a formal statement of its inputs, outputs, and evaluation under the prevalence at which documentation updates occur.}
  \item \revision{A proof-of-concept framework for the task, evaluated against a single-prompt baseline that isolates the contribution of its staged design.}
  \item A comprehensive failure case analysis that characterises the nature of prediction errors and outlines directions for future system enhancement.
  \item A retrospective case study demonstrating the tool's capability to identify overlooked documentation updates, complemented by an assessment of developer perspectives. 
\end{itemize}


Collectively, these contributions advance documentation maintenance by shifting the focus from element-wise identification or entire regeneration to surgical, human-in-the-loop updates within pull-based development activities. \revision{This work gives researchers a task formulation and an evaluation protocol for documentation maintenance in pull-based development, together with a framework and a set of measurements that establish where the task currently stands.}

\revision{\section{Motivating Example}}

\revision{README files often function as trusted operational guidance. Users rely on them to install software, configure authentication, choose among third-party tools, and onboard onto a project. The severity differs from case to case, but outdated content can impose real costs on everyone who depends on it. In this section we present one such case to motivate our study.}

\revision{The case concerns \texttt{youtube-dl}, an open-source command-line tool for downloading online videos. At the time of the incident its latest stable release was version 2021.12.17, which had been published more than a year earlier. The FAQ in its README nevertheless told users to update the tool by running \texttt{youtube-dl -U}, or \texttt{sudo youtube-dl -U} on Linux. In February 2023 a change to the metadata served by YouTube caused the tool to fail with an \texttt{Unable to extract uploader id} error. The defect was fixed on the main branch the following day, but no new stable release followed, so the documented update command still resolved to the frozen 2021.12.17 build and reported that the user was already up to date. The working instructions lived in \href{https://github.com/ytdl-org/youtube-dl/issues/31530}{issue~\#31530}, where maintainers noted that \texttt{-U} ``doesn't (yet) work'' and directed users to install from the main branch, and later a nightly build. Of the 670 issues opened in the three and a half months after the breakage, 518 (77\%) concerned this same already-fixed failure.}

\revision{This inconsistency imposed costs on both users and maintainers. Users continued to submit duplicate reports long after a fix was available. The maintainers had to redirect them to the issue discussion and warn that further duplicate comments could be removed or marked as spam. Some users also asked whether they should move to the actively released \texttt{yt-dlp} fork. The effects later extended into the packaging ecosystem. The Homebrew project lead asked the maintainers for any tagged release so that downstream users would stop filing these reports, and when none appeared Homebrew deprecated the package in November 2023, citing \href{https://github.com/ytdl-org/youtube-dl/issues/31585}{issue~\#31585} and the absence of a new release in the formula itself. It disabled the formula in November 2024 and removed it in November 2025. The Homebrew decision reflected broader maintenance concerns and cannot be attributed to the README alone. Even so, this case shows how outdated update instructions can obscure an available fix, prolong user-facing failures, increase support effort, and contribute to migration pressure. A documentation check triggered by the PR carrying the fix could have surfaced this mismatch immediately upon entering the main branch, prompting maintainers to revise the FAQ}

\revision{This example illustrates the practical impact of a poorly maintained README. It motivates both the need to keep documentation up to date and the development of tools that recommend timely updates to maintainers. If the README had reflected the availability of the fix and the correct installation procedure, users could have resolved the problem without relying on a lengthy issue discussion. This would have reduced duplicate reports, repeated replies, and the ongoing effort required to moderate the issue thread.  More broadly, the case demonstrates how recommending a README update upon introducing a relevant code change can reduce avoidable communication overhead for project maintainers.}

\section{Related Work}

In this section, we discuss software documentation issues and automated solutions for documentation maintenance.

\subsection{Software Documentation Issues}

Software documentation is a key artefact among stakeholders. Garousi et al.~\cite{garousi2015usage} conducted an industrial case study on the usage and utility of various documentation artefacts throughout the software development lifecycle, revealing that documentation needs vary significantly across development stages. These documentation categories include code comments~\cite{fluri2007code}, API documentation~\cite{maalej2013patterns}, among other documentation artefacts.

README file, the primary artefact examined in this study, serves as an overview of the underlying repository. Prana et al.~\cite{prana2019categorizing} investigated categorisation of the content of README files. They observed that README files primarily contain sections addressing ``What'' the software is and ``How'' to use it, and subsequently developed an automated approach to classify these contents. Furthermore, Liu et al.~\cite{liu2022readme} examined the association between repository popularity and README structures in Java projects, identifying a specific structural organisation associated with higher popularity. Most recently, Gao et al.~\cite{gao2025adapting} conducted a large-scale empirical study on README update patterns to establish a comprehensive taxonomy of updates and their corresponding triggering events. Furthermore, their proposed template for modularising README content, aimed at improving presentation quality and reducing maintenance effort~\cite{treude2020beyond}, has been endorsed by active OSS repositories.

\revision{However,} software documentation suffers from quality degradation. Research indicates that stakeholders prioritise content integrity over presentation. Uddin and Robillard~\cite{uddin2015api} found that practitioners view content deficiencies in API documentation as more problematic than presentation flaws. Among these content issues, documentation outdatedness emerges as a predominant concern. Large-scale mining studies of developer discussions on GitHub and Stack Overflow identify outdated documentation as a recurrent theme~\cite{aghajani2019software}, a finding further validated by user-centric surveys where developers rank it as a top-tier challenge~\cite{aghajani2020software}. This struggle is pervasive, extending even to the emerging domain of AI-based system development, where incomplete and outdated information continues to impede progress~\cite{banyongrakkul2025from}.

\subsection{Automated Documentation Generation and Augmentation}

Automated solutions have been investigated to enhance documentation through either generation or augmentation. For example, generative techniques have been widely applied to produce code comments~\cite{fowkes2017autofolding}, commit messages~\cite{xu2019commit}, and release notes~\cite{moreno2014automatic}. These approaches assist developers in creating such artefacts from scratch, thereby mitigating the issue of missing documentation. Furthermore, augmentation approaches aim to improve the quality and usability of existing documentation. Stylos et al.~\cite{stylos2009improving} enhanced API documentation by systematically incorporating usage examples. Similarly, Treude and Robillard~\cite{treude2016augmenting} proposed a method to augment API documentation by retrieving relevant insights from Stack Overflow. More recently, Gao et al.~\cite{gao2023evaluating} adapted transfer learning to simplify technical jargon in README files, thereby enhancing comprehension for entry-level contributors.

Regarding the up-to-dateness issue, the rapid evolution of codebases necessitates active maintenance to ensure that software documentation remains current. Dagenais and Robillard~\cite{dagenais2014using} identified coherent code-like elements within documentation, leveraging them as traceability links to codebase elements to infer outdated API segments. Building on this, Tan et al.~\cite{tan2024detecting} proposed an approach for detecting mismatches between code snippets in README files and the codebase. However, while such rule-based approaches effectively validate specific code literals, they are inherently limited by their reliance on exact matching, rendering them unable to reason about outdated information expressed in natural language or within abstract artefacts like READMEs.

The rapid advancement of LLMs has enabled new approaches to documentation maintenance. Gao et al.~\cite{gao2025adapting} proposed a template of common README sections, leveraging LLMs to rephrase content for clearer, modularised updates. This is an initial effort to facilitate document maintenance, but cannot detect outdated components. Parallel research has explored generating project-level documentation from scratch; for example, Cui et al.~\cite{11186105} designed an agentic system that employs iterative tool invocations to generate full README files. However, regenerating the entire document upon every development activity is computationally expensive and imposes a significant review burden on developers~\cite{sadowski2018modern}, while also introducing risks of stylistic drift~\cite{geiger2018types} and incorporation of hallucinated content~\cite{liu2024lost}. Focusing on maintaining existing artefacts, Lee et al.~\cite{Lee2025API} applied LLMs to revise API documentation. Yet, their assessment relies heavily on automated metrics like BLEU~\cite{papineni2002bleu}, which are recognised as unsuitable proxies for quality in software documentation tasks due to their poor correlation with human judgement~\cite{hu2022correlating}. Furthermore, their detection of outdated components remains constrained by a reliance on rule-based matching of code snippets.

Existing approaches fail to comprehensively resolve the issue of outdated README files, primarily due to their inability to detect high-level semantic mismatches. This limitation forces a continued reliance on developers to manually verify documentation status, a fragile strategy given that documentation is often neglected~\cite{forward2002relevance}. Furthermore, full-scale generation methods remain prohibitively expensive for continuous integration as they necessitate scanning the entire codebase for every incremental change, subsequently imposing a significant review burden on developers. To address these challenges, we propose an LLM-driven approach that selectively utilises granular information from PR to recommend targeted updates. This strategy not only minimises the cognitive load and review time for maintainers but also integrates documentation maintenance directly into the development workflow.

\section{Description of Dataset and Approaches}
\label{sec:approach}

In this section, we first provide an example to outline the expected inputs and outputs of our approach using a real-world scenario (Section~\ref{sec:motivating_example}).
We then describe the dataset constructed for our study (Section~\ref{sec:dataset}). Subsequently, we detail the setup of our automated approach for README updates. We begin by introducing the core components that serve as the fundamental building blocks of our system (Section~\ref{sec:core_components}). We then describe the orchestration strategies, including a static version (Section~\ref{sec:workflow}) and an agentic version (Section~\ref{sec:agentic_design}), to demonstrate how these components interact to handle PRs and recommend documentation updates.

\subsection{Task Formulation}

{\revisioncolor
We state the task that our framework instantiates. Let $p$ denote a PR and let $D$ denote the state of a documentation artefact in the repository at the time $p$ is opened. We parse $D$ into an ordered sequence of atomic units
\begin{equation}
  \sigma(D) = \langle u_1, u_2, \ldots, u_n \rangle,
\end{equation}
where each $u_i$ is the smallest unit at which a maintainer would apply an edit. For a README written in Markdown, we take this unit to be a paragraph, and we index the units in document order, as described in Section~\ref{sec:rq1-methodology}.

The task can be abstracted to a function
\begin{equation}
  f: (p, D) \mapsto \bigl(y, \, L, \, J \bigr),
\end{equation}
where $y \in \{0, 1\}$ indicates whether $p$ invalidates any part of $D$, $L = \langle i_1, \ldots, i_m \rangle$ with $m \leq k$ is a list of unit indices ranked by the estimated likelihood that the corresponding unit requires revision, and $J = \langle j_1, \ldots, j_m \rangle$ is a corresponding list of natural language justifications, each linking $u_{i_t}$ to the element of $p$ that invalidated it. When $y = 0$, the lists $L$ and $J$ are empty. The justifications $J$ inform the maintainer's decision and keep the human in the loop. 

Two properties of the task follow from its trigger.
Because $f$ is invoked once per PR, the marginal distribution of $y$ is that of real development activity, in which documentation updates are rare. A configuration of $f$ is therefore characterised by a pair of conditional rates rather than by a single accuracy figure, and the quantity of interest to a maintainer is the posterior probability that a surfaced recommendation is correct under that prevalence. Because the output is a set of indices into an existing document, the correctness of $L$ is graded rather than binary, and a recommendation that identifies the correct region of the document is more useful than one that does not, even when it misses the exact unit. Section~\ref{sec:rq1-methodology} defines the metrics that follow from these two properties.

Supervision for $f$ is available only in proxy form.
The signal observable at scale is the set of units that developers did edit in the same PR, which records what they did rather than what the document required. These two disagree in both directions. Some recorded edits are cosmetic, or respond to earlier changes rather than to the PR at hand, which we examine in Section~\ref{sec:rq2b}. Some PRs carry no recorded edit even though the README had become inaccurate, which we quantify in Section~\ref{sec:rq3}. We adopt this record as our empirical ground truth throughout, and results measured against it therefore express agreement with historical maintenance behaviour.
\par}

\revision{\subsection{Pipeline Input and Output}}\label{sec:motivating_example}

\begin{figure}[t!]
    \centering
    \includegraphics[width=1\linewidth]{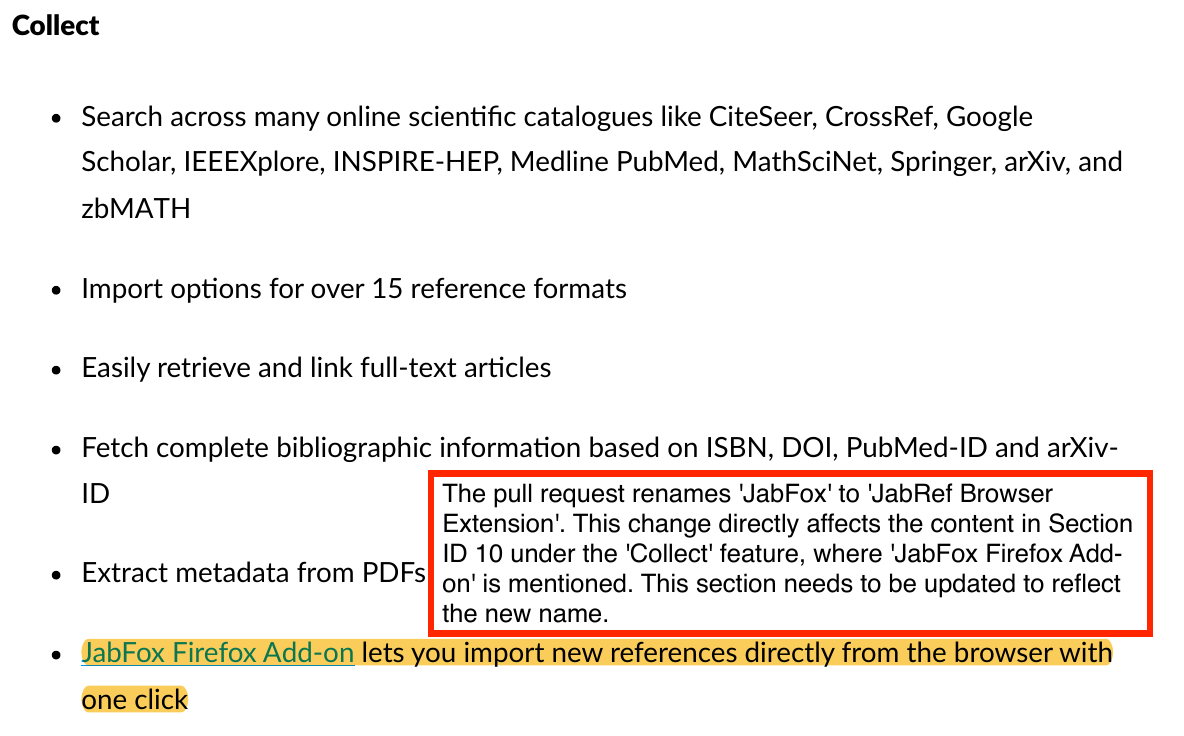}
    \caption{\revision{Sample Output Example}}
    \label{fig:motivating_example}
\end{figure}

To provide a high-level overview of our approach, we present a real-world scenario drawn from our in-person evaluation with \texttt{jabref}. Figure~\ref{fig:motivating_example} illustrates this demonstration.

Upon receiving PR \textit{\#5575}, our approach analyses the PR context, including PR description, commit messages, along with relevant file patches, to reveal that the component ``JabFox'' has been updated to the new name ``JabRef Browser Extension'' to enable more extensions other than Firefox. Subsequently, the approach isolates the specific section within the README file that has become outdated (highlighted in orange), supplying a justification detailing why this modification is necessary (indicated by the red box).

\subsection{Data Preparation and Dataset Characteristics}\label{sec:dataset}

As our automated approach relies on understanding and evaluating real-world update patterns, we require a historical dataset for evaluation. Therefore, we focus our investigation on past documentation updates and their corresponding code changes at the PR level. Specifically, we utilise PRs that contain modified README files as our ground truth. To construct this dataset, we leveraged the collection from Lin et al.~\cite{lin2025leveraging}, which contains the complete PR histories of 826 top GitHub repositories. This corpus encompasses 4,423,627 PRs across projects in nine popular languages, including Python, Java, Go, C++, JavaScript, C, C\#, PHP, and Ruby. However, it is important to note that ground truth reflects observed developer behaviour, not necessarily optimal documentation quality.

Since our primary focus is on README updates triggered by codebase changes, we initially filtered the dataset to include only PRs containing at least one modification to the root README alongside changes to other files. This refinement yielded 35,482 PRs across 780 repositories. Notably, while 94.4\% of the repositories in the dataset contain at least one PR with a README update, such updates are rare in the broader context, with only 0.8\% of all PRs in the dataset involving modifications to the README file.

To further isolate updates driven by code evolution, we applied stringent filtering criteria. We removed PRs containing the keyword ``README'' in their title, as well as those where the README update did not chronologically follow a codebase modification, applying a five-minute threshold to account for quick fixes~\cite{wen2020empirical}. These steps effectively remove PRs where the primary intention was solely to update documentation, which are often triggered by procedural project decisions rather than code evolution. After applying these filters and excluding repositories no longer accessible on GitHub, our final dataset comprises 15,190 PRs across 714 repositories.

We applied filters to exclude outliers, specifically overly-lengthy README edits and PRs containing an excessive number of commits or file changes, based on the intuition that extremely large edits are atypical of routine documentation maintenance and could skew the evaluation. Similarly, oversized PRs often reflect poor development practice, failing to represent a coherent, single unit of development activity~\cite{herzig2013impact}. We determined these thresholds using Tukey's fences~\cite{tukey1993exploratory}, resulting in the exclusion of PRs exceeding 11 edited paragraphs in the README, 145 changed files, or 23 commits.

We constructed two datasets: a positive set comprising \revision{10,702} PRs that updated a README, and a negative set of 14,809 PRs randomly sampled from those where a README was not updated. The numbers both drop from 15,190 as we originally sampled the same amount and further applied the same filtering criteria based on the outlier thresholds. We refer to these as the ``positive'' and ``negative'' datasets throughout the remainder of this paper. Each sample in the dataset contains the following items: (1) repository name, (2) PR number, (3) README file content before any update is applied, (4) a list of commits within the PR, (5) PR title and description, (6) a list of changed files and their patches within the PR, and (7), a README file patch (if the PR contains a README update).

\begin{figure*}[!t]
    \centering
    \includegraphics[width=\linewidth]{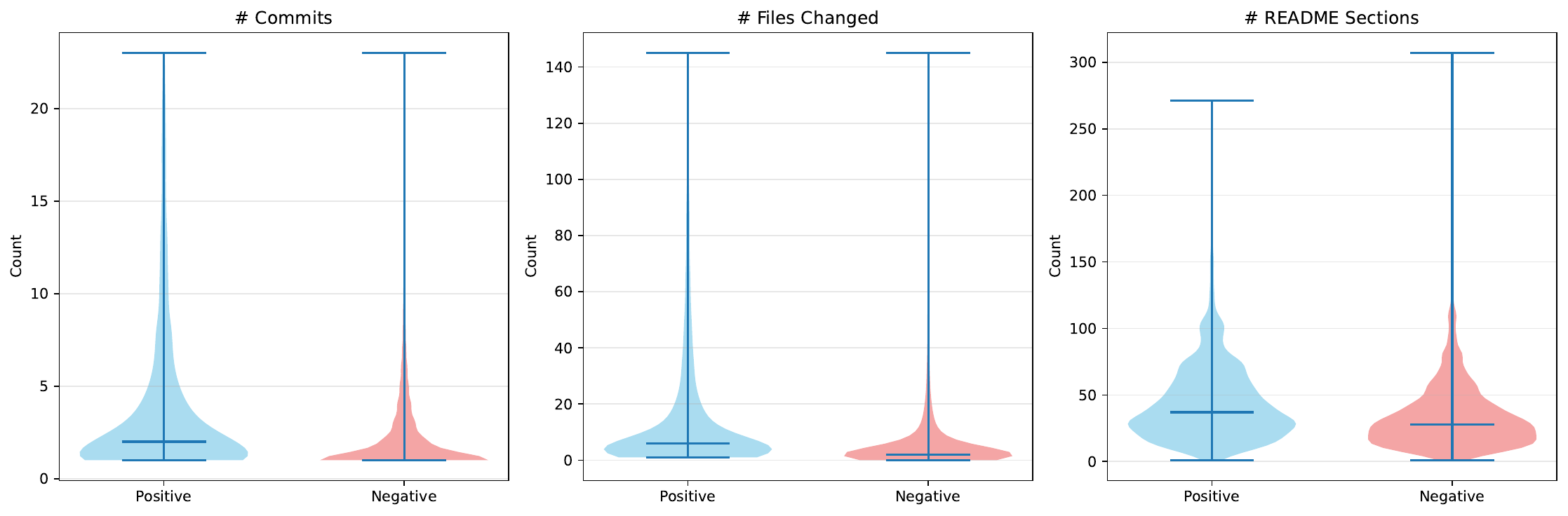}
    \caption{Comparison of positive and negative PRs in our dataset.}
    \label{fig:data-characteristics}
\end{figure*}

Figure~\ref{fig:data-characteristics} presents violin plots comparing the distributions of commit counts, updated file counts, and pre-update README section counts between positive and negative PRs. The data indicates that PRs involving README updates exhibit greater complexity, characterised by a higher volume of commits and modified files. On average, positive samples contain 44.5 README sections. Consequently, the model must possess strong reasoning capabilities to align the PR's intention with the existing document structure. Furthermore, with an average of 16.6 modified files per positive PR, it is impractical to feed the entire context into the model. This constraint validates our architectural choice of a pipeline specifically engineered to prioritise the most relevant information.

\subsection{Core Components}\label{sec:core_components}

Developing a reliable automated approach requires addressing the severe class imbalance inherent to documentation maintenance. As established in Section~\ref{sec:dataset}, README updates occur in less than 1\% of PRs. Our initial exploratory experiments on a small negative subset, which required the LLM to perform an end-to-end evaluation using all available PR context information, yielded highly ineffective results. Specifically, this naive approach produced a false positive rate exceeding 60\%, not useful for providing useful feedback.

\begin{figure*}[!t]
    \centering
    \includegraphics[width=\linewidth]{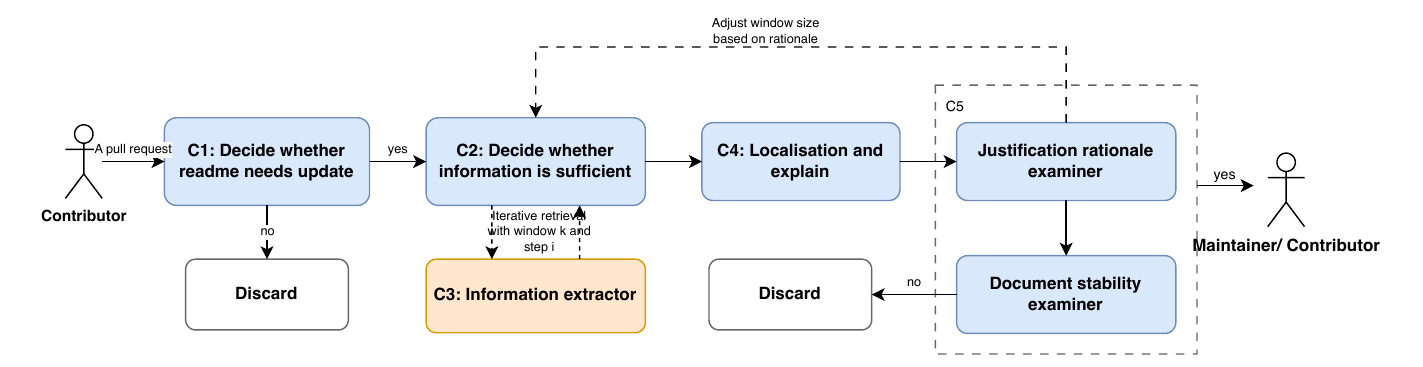}
    \caption{Design of the Document Update Recommendation Pipeline.}
    \label{fig:design-framework}
\end{figure*}

To overcome these limitations, we designed a multi-stage pipeline that decomposes the complex maintenance task into distinct, verifiable steps: aggressively filtering irrelevant PRs, dynamically evaluating and retrieving necessary context, generating targeted updates, and strictly reviewing the output for quality assurance. Furthermore, to determine the optimal way to connect and execute these steps, we propose and evaluate two different orchestration strategies: a linear static workflow (Section~\ref{sec:workflow}) and an iterative agentic workflow (Section~\ref{sec:agentic_design}). Our system is composed of five distinct components, each designed to execute a specific sub-task within the documentation maintenance workflow. Figure~\ref{fig:design-framework} illustrates the overall architecture and the interaction between these components. The individual components are discussed further in Section~\ref{sec:component_analysis}, while the prompts used are accessible in the replication package. \revision{Throughout, a component denotes a unit that takes a defined input, performs one step of reasoning or retrieval, and returns a defined output, realised either as a single round of LLM prompting or as a deterministic retrieval procedure that makes no model call. The decision rules that act on those outputs, and the bounds on any repetition, are properties of the orchestration rather than of a component, which is what allows the static and agentic workflows to be compared over the same five components.} \revision{The components differ in how they are realised. The components in blue represent distinct LLM prompts, while the component in yellow represents a retrieval heuristic. The arrows between components are decision rules based on LLM output, with the dashed arrows indicate autonomous behaviour that can be conducted in the agentic version. Each component is thus specified by its prompt, its decoding configuration and the decision it returns, which we detail in Section~\ref{sec_experimental_configuration} and release in the replication package. Section~\ref{sec:component_analysis} then reports how each component behaves when it is reconfigured in isolation.}

\begin{itemize}
    \item \textbf{C1 (Relevance Classifier):} This component acts as the initial gatekeeper. It receives the PR contexts and README file as input to determine whether the PR necessitates a README update. Its primary objective is to filter out irrelevant PRs at an early stage, thereby optimising computational efficiency by discarding candidates that clearly do not require modification.
    \item \textbf{C2 (Context Sufficiency Analyser:)} This component evaluates the information density of the inputs. It assesses whether the available context (initially comprising the PR title, description, and commit messages) provides sufficient evidence to confidently localise the required update and formulate a justification. It outputs a binary decision: \textit{Sufficient} or \textit{Insufficient}.
    \item \textbf{C3 (Context Retrieval Unit:)} This component is responsible for acquiring auxiliary information to bridge context gaps. Specifically, it retrieves file patches from the set of files modified in the PR. It employs a similarity-based metric to rank these patches, selecting the most relevant segments to augment the context for downstream reasoning components.
    \item \textbf{C4 (Localisation and Generation Engine:)} Serving as the central reasoning unit, this component leverages the accumulated context to perform two critical tasks: (1) it identifies the specific section indices in the README that require modification, and (2), it synthesises a natural language justification explaining \textit{why} the update is necessary relative to the underlying code changes.
    \item \textbf{C5 (Quality Assurance (QA) Reviewer):}  To mitigate false positives and hallucinations, C5 operates as a dual-purpose critic. It validates the quality of the generated justification, ensuring the rationale is specific, actionable, and aligned with the README's existing granularity. Simultaneously, it assesses documentation stability, verifying that the proposed update is strictly necessary given the current document status.
\end{itemize}

\subsection{Static Workflow}\label{sec:workflow}
The static workflow, denoted by the solid arrows in Figure~\ref{fig:design-framework}, executes the components in a strict linear sequence. This configuration serves as our baseline. The process initiates when a PR is created. The workflow proceeds as follows: 

\begin{enumerate} 
\item \textbf{Gatekeeping:} C1 performs an initial assessment by evaluating the PR metadata (title, description, and commit messages) against the current README content. If the PR is classified as ``No Update Required'', the workflow terminates to conserve computational resources. 
\item \textbf{Context Acquisition:} C2 evaluates the informational sufficiency of the initial PR context. If the context is deemed insufficient, C3 is triggered to retrieve a fixed quantity of auxiliary information (e.g., the top-1 ranked file patch) to augment the input. Conversely, if the context is sufficient, the workflow advances directly to the generation phase.
\item \textbf{Generation:} C4 leverages the accumulated context, comprising the original PR metadata and any retrieved patches, to identify the target section indices and synthesise a justification for the proposed update.
\item \textbf{Final Review:} C5 executes a conclusive single-pass review. It validates the recommendation, approving it only if the justification satisfies the predefined specificity and stability criteria. Should the validation fail, the recommendation is discarded.
\end{enumerate}

\subsection{Agentic Design}\label{sec:agentic_design}

While straightforward, the static workflow is rigid. It enforces an identical execution path for every PR regardless of the complexity, preventing the system from recovering if the initial retrieval (C3) fetches irrelevant files or if the justification (C4) is overly generic. To address the inherent rigidity of the static version, we introduced an agentic workflow that incorporates dynamic feedback loops. We explored this option as agentic solution has demonstrated better performance in similar tasks, such as commit message generation~\cite{li2024only}. We followed existing guidelines for agentic system design~\cite{cai2025designing, shinn2023reflexion} with self-reflection to iteratively reason about the optimal approach for capturing the PR intention and facilitating accurate README updates.

We first implemented a Dynamic Information Retrieval Loop ($C2 \leftrightarrow C3$). Unlike the static baseline, if C2 deems the context insufficient, it triggers C3 to iteratively retrieve the \textit{next} most relevant file patch. This cycle repeats up to $p$ times or until sufficiency is met, allowing the system to reason over the most informative context available within the PR.

Furthermore, we incorporated a Self-Reflecting Refinement Loop ($C5 \rightarrow C2$) to validate the reliability of the intermediate reasoning steps. Specifically, the QA Reviewer (C5) critiques the generated justification to diagnose the quality of the underlying context. If the justification is deemed ``Overly Generic'', the system infers an information deficit and consequently expands the retrieval window to ingest richer context. Conversely, if the justification is flagged as ``Hallucinated'', referencing non-existent files, code, or documentation segments, the system hypothesises that the model is overwhelmed by irrelevant noise (the ``Lost in the Middle'' phenomenon). In this scenario, it contracts the retrieval window to sharpen the signal. This bidirectional adjustment ensures the model dynamically converges on the optimal context size, effectively balancing sufficiency with clarity.

\revision{
\subsection{Experimental Configuration and Prompt Design}
\label{sec_experimental_configuration}

All components are served with vLLM and process one pull request at a time, without batching across pull requests. We set the temperature to 0 in every component so that a given input yields the same output on repeated runs. Components that emit a binary decision, namely the Relevance Classifier (C1) and the Quality Assurance Reviewer (C5), together with their counterparts in the agentic workflow, use constrained decoding restricted to the two tokens \texttt{Yes} and \texttt{No}, generate a single token. Components that produce structured text, namely the Sufficiency Analyser (C2), the Context Retrieval Unit (C3) and the Localisation and Generation Engine (C4), decode freely with a limit of 1{,}024 output tokens and return a JSON object that we parse into a list of section identifiers and a justification. The context window is 32{,}768 tokens for every model. Inputs that exceed the window are truncated. Prompts are rendered with each model's chat template with the generation prefix enabled and, for the models that expose a reasoning mode, with that mode disabled, so that every component returns a direct answer rather than an intermediate reasoning trace. The larger models are served on two NVIDIA A100 80\,GB GPUs and the smaller models on one.

We developed the prompts iteratively on a working sample of 500 PRs equally distributed on the positive and negative samples. Rather than asking a single model to perform the complete task in one instruction, we designed a separate prompt around the responsibility of each component. Each prompt states the component's task, presents the available inputs in consistently labelled fields, defines the criteria governing its decision, and specifies the required output format. During development, we evaluated the intermediate decisions on this sample and revised the corresponding decision criteria when recurrent errors were observed. We repeated this process until the prompts produced stable performance, and then used the resulting prompts in the reported experiments. This sample was not held out, but prompt development revised the general decision criteria of each component rather than fitting any parameter to individual PRs, and the sample accounts for only 2.3\% of the positive and 1.7\% of the negative evaluation set.
}

\section{Research Questions}
\label{sec:RQs}
In the following we refer to our approach as ``the pipeline'' since it follows a pipeline-based design (see Section~\ref{sec:approach}). To systematically evaluate the performance and utility of our proposed pipeline, we structure our investigation around the following three research questions:

\begin{tcolorbox}[left=1pt, top=1pt, right=1pt, bottom=1pt] 
\noindent \textbf{RQ1 (End-to-End Evaluation):} 
\begin{itemize}
\item \textbf{RQ1a:} How accurately can the pipeline flag PRs that require a README update?
    \item \textbf{RQ1b:} How precisely can the pipeline localise the sections within the README that need modification?
\end{itemize}

\noindent \textbf{RQ2 (Component \& Failure Analysis):} 
\begin{itemize}
    \item \textbf{RQ2a:} How do the pipeline's component designs and reasoning strategies address the challenges of context selection and severe class imbalance?
    \item \textbf{RQ2b:} What are the characteristic failure modes that persist in the pipeline?
\end{itemize}
    
\noindent \textbf{RQ3 (Identification of Overlooked Updates):} To what extent can the pipeline identify previously-overlooked README updates?
\end{tcolorbox}

\begin{figure}
    \centering
    \includegraphics[width=\linewidth]{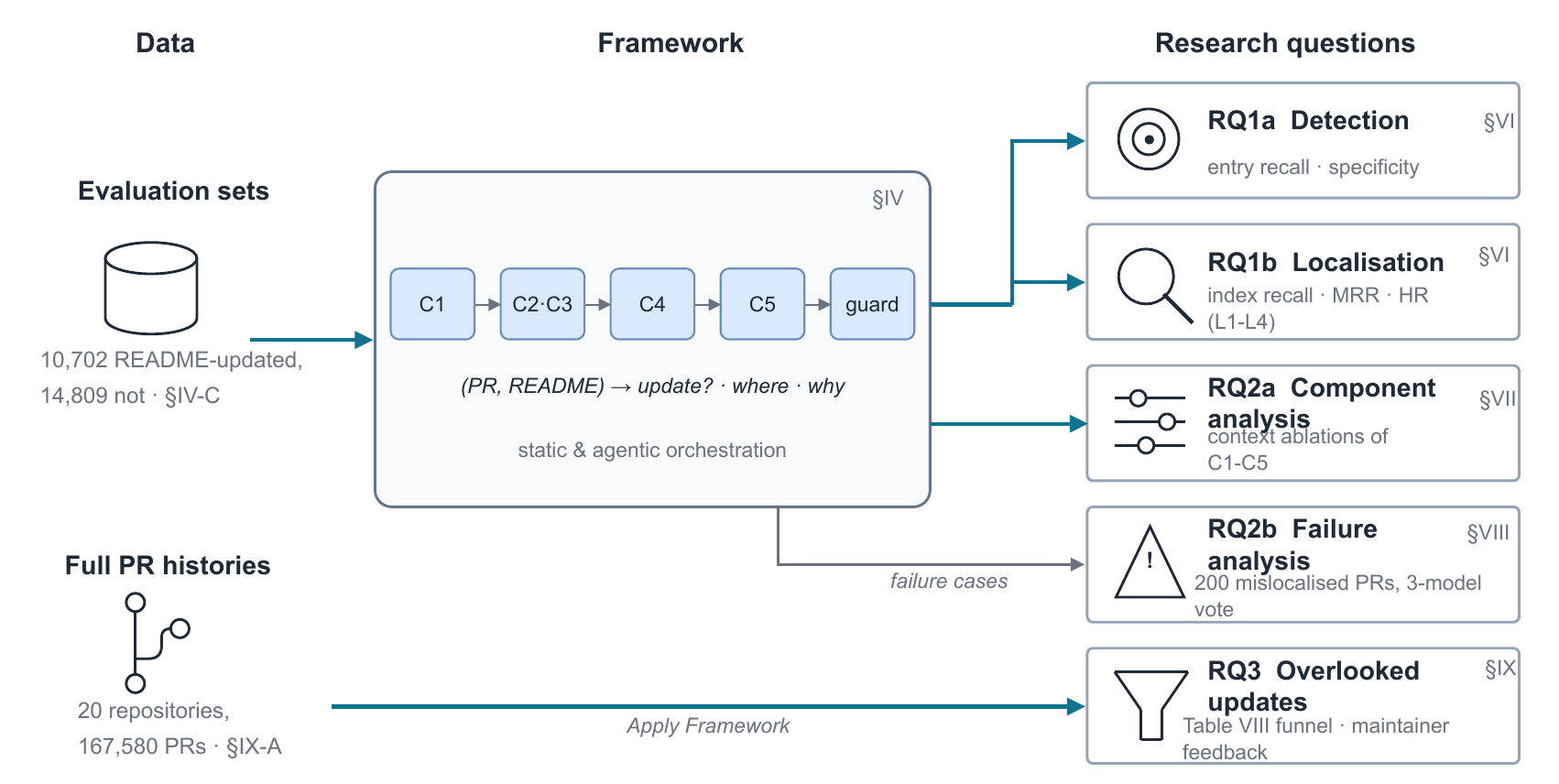}
    \caption{Study Overview}
    \label{fig:overview}
\end{figure}

\revision{Figure~\ref{fig:overview} provides the study overview.} To address RQ1, we conducted an end-to-end evaluation to assess the holistic effectiveness of the approach in accurately detecting update requirements and localising outdated documentation sections. For RQ2, we deconstructed the pipeline, validating the contribution of individual components through ablation studies, manual annotation, and qualitative analysis. Finally, to answer RQ3, we perform a retrospective study across the complete PR history of 20 sampled repositories, complemented by a targeted case study on a specific OSS project, to evaluate real-world applicability and developer reception.


\section{End-to-End Evaluation (RQ1)}

The end-to-end evaluation assesses performance across two primary dimensions: first, the capability to correctly flag PRs that necessitate a README update, and second, the localisation accuracy of the recommended updates for PRs containing ground truth documentation changes.

\subsection{Methodology} 
\label{sec:rq1-methodology}

We first evaluate the system's effectiveness in distinguishing between PRs that require updates and those that do not. While this corresponds to a standard binary classification task, we employ specific terminology to distinguish these PR-level predictions from the subsequent granular localisation tasks. We assess performance using three metrics, defined based on the standard confusion matrix of True Positives ($TP$), False Negatives ($FN$), True Negatives ($TN$), and False Positives ($FP$):
\begin{itemize}
  \item \textbf{Entry Recall:} The proportion of PRs \revision{that were accompanied by a README update in the historical record} that are correctly identified by the approach:

\begin{equation} \frac{TP}{TP + FN} \end{equation}
\item \textbf{Entry Specificity:} The proportion of PRs \revision{that were not accompanied by a README update} that are correctly filtered out by the approach:
    \begin{equation}
    \frac{TN}{TN + FP}
    \end{equation}
    \item \textbf{User-Facing Accuracy:} The percentage of generated recommendations presented to the user that are actually correct. \begin{equation} \frac{\text{Recall} \times 1}{\text{Recall} \times 1 + (1 - \text{Specificity}) \times 99} \end{equation}
\end{itemize} 
All metrics are benchmarked against the actual update behaviours of developers observed in the dataset, which serves as our empirical ground truth. \revision{This record is a proxy rather than an oracle of documentation necessity, since a README edit accompanying a PR may be cosmetic, unrelated or bundled with other cleanup, and a PR without one may still have warranted an update, so all metrics reported here measure agreement with what developers did rather than with what was necessary.} Given that the distribution of PRs requiring README updates is highly skewed in practice, reporting raw precision on a balanced test set would be misleading. Therefore, we calculate this metric based on a simulated real-world prevalence ratio of \textbf{1:99} (assuming only 1\% of PRs necessitate a README update). This ratio is directly supported by our corpus analysis in Section~\ref{sec:dataset}, which established that README updates empirically occur in only 0.8\% of the 4.4 million PRs analysed. This adjustment ensures the metric accurately reflects the signal-to-noise ratio a developer would experience in a production environment, where high false positive rates are a known primary driver of tool abandonment~\cite{johnson2013don}.

We then evaluate the localisation accuracy of the generated indices for documentation updates. Our approach splits the README file into separate sections based on line changes, which corresponds to paragraph-level granularity in Markdown syntax. Each section is assigned with an index, starting from one. Beyond merely flagging the need for an update, our system identifies specific locations (indices) within the document where updates are required, alongside a justification for each (see Section~\ref{sec:motivating_example}). These output indices are ranked by their predicted probability of requiring modification.

\begin{table*}[t]
\revisioncolor
\centering
\caption{End-to-end comparison between fixed pipeline and agentic versions. Hierarchical Recall is reported as a tuple (L1, L2, L3, L4).}
\label{tab:end-to-end}
\begin{tabular}{l | r r r r | r r r }
\toprule
Model & Version & Entry Recall & Entry Specificity & User-facing Accuracy & Index Recall & MRR & Hierarchical Recall \\
\midrule
\multirow{2}{*}{Gemma-3-27b} & Static & 0.53 & 0.90 & 0.05 & 0.60 & 0.66 & (0.95, 0.82, 0.68, 0.70) \\
 & Agentic & 0.69 & 0.94 & 0.10 & 0.61 & 0.66 & (0.95, 0.83, 0.71, 0.73) \\
\midrule
\multirow{2}{*}{Qwen-3-32b} & Static & 0.49 & 0.88 & 0.04 & 0.63 & 0.68 & (0.96, 0.87, 0.77, 0.77) \\
 & Agentic & 0.50 & 0.99 & 0.28 & 0.68 & 0.73 & (0.97, 0.90, 0.83, 0.81) \\
\midrule
\multirow{2}{*}{Llama-3.3-70b} & Static & 0.38 & 0.94 & 0.07 & 0.68 & 0.72 & (0.97, 0.86, 0.75, 0.76) \\
 & Agentic & 0.54 & 0.89 & 0.06 & 0.66 & 0.71 & (0.96, 0.86, 0.75, 0.73) \\
\midrule
Qwen-3-32b & Baseline prompt & 0.77 & 0.70 & 0.03 & 0.56 & 0.61 & (0.95, 0.85, 0.70, 0.69) \\
Random Guesser & ------ & 0.01 & 0.99 & 0.01 & 0.11 & ------ & ------ \\
\bottomrule
\end{tabular}
\end{table*}

Drawing on fault localisation methodologies~\cite{xu2025flexfl}, we explicitly instructed the LLM to return the top-$k$ candidate indices, setting $k=5$. We selected this threshold for two primary reasons. First, prior research indicates that 73.68\% of developers only inspect the top five suggested elements~\cite{kochhar2016practitioners}. Second, this value aligns with the empirical distribution of ground truth updates in our dataset, where the average number of edited sections is 1.7, and 96.8\% of updates involve five or fewer paragraphs. We assess the accuracy of these recommended indices using three metrics:

\smallskip\noindent\textbf{Index Recall:} Recall evaluates the percentage of ground truth indices correctly predicted by our approach, using the formula $\frac{|G\cap P|}{|G|}$. Here, $G$ and $P$ denote the sets of indices in the ground truth and the prediction, respectively.

\smallskip\noindent\textbf{Mean Reciprocal Rank (MRR):} For the predicted section indices, we require the model to output a list of indices ranked by the likelihood that the corresponding section requires an update. MRR evaluates the quality of this ranking by identifying the position of the first correct prediction. Intuitively, a higher MRR indicates that the correct target section appears earlier in the recommendation list. The metric is calculated as follows:
    \begin{equation}
        \text{MRR} = \frac{1}{|Q|} \sum_{i=1}^{|Q|} \frac{1}{\text{rank}_i}
    \end{equation}
    \begin{equation}
        \text{rank}_i = \min \{ k \mid P_{i,k} \in G_i \}
    \end{equation}
    
\noindent $|Q|$ denotes the total number of entries in the positive dataset, and $P_{i,k}$ is the $k$-th item in the predicted list for query $i$. $G_i$ is the set of ground truth indices for query $i$.

\smallskip\noindent\textbf{Hierarchical Recall:} The preceding metrics do not account for semantic proximity. They treat all incorrect predictions as equivalent errors, assigning a score of zero regardless of structural context. However, two indices residing under the same header are semantically closer than indices located in disparate headers\footnote{For instance, two paragraphs within the ``Installation'' section are both relevant to the installation process and are, thus, semantically closer than a paragraph in the ``Contribution Guidelines''.}. Intuitively, a recommendation that lands within the correct parent section, even if not on the exact paragraph, helps developers localise the issue more effectively than a completely irrelevant suggestion. Therefore, we aim to measure this granularity by using a hierarchical score.
    
To implement this, we parse the README file into a hierarchical tree structure determined by header levels. In this schema, Level 1 headers serve as top-level nodes, with subsequent sub-headers nested as children. We recursively construct the tree to a maximum depth of Level 4; any content appearing subordinate to a Level 4 header is aggregated into a single leaf node without further segmentation. This depth restriction is based on prior research~\cite{gao2025adapting}, which identified that the top four header levels capture the primary structural information, whereas lower-level headers are predominantly used for text highlighting or formatting purposes. 

To calculate hierarchical recall at level $i$, we employ a relaxed matching criterion: a predicted index is considered accurate if it resides within the same node at level $i$ as a ground truth index. The metric for level $i$ is calculated as: 
\begin{equation}
    Recall_i = \frac{|N_i(G)\cap N_i(P)|}{|N_i(G)|}
\end{equation}
\noindent Where $N_i(S)$ denotes the set of unique nodes at level $i$ corresponding to the indices in set $S$.

Within our positive PR dataset, README files exhibit a hierarchical structure comprising an average of 2.30 Level-1 (L1), 6.10 L2, 3.15 L3, and 1.03 L4 section headers. In contrast, the ground truth edits demonstrate high specificity, impacting an average of only 0.85, 0.88, 0.31, and 0.09 of these respective sections per update. We conducted all experiments with three recently released open-weight LLMs that have been used in past research: Llama-3.3-70B, Qwen-3-32B, and Gemma-3-27B. We set the temperature to 0 for all models in all experiments to minimise the randomness.

\subsection{Results} 


Table~\ref{tab:end-to-end} presents the end-to-end evaluation results, comparing the performance of the static pipeline against the agentic pipeline. \revision{The three localisation metrics are computed to reflect the final delivered recommendation. Components of the metrics (i.e., TP/FP/TN/FN) are recorded in Appendix~\ref{append-component}.} To contextualise these findings within the severe 1:99 class imbalance, we established a weighted random baseline as a point of comparison, \revision{representing the behaviour of neglecting documentation updates altogether}. For each PR, this baseline assigns a positive label with a probability of 0.01 and randomly selects five candidate indices for updates. As shown in the results, this naive approach yields a user-facing accuracy of only 0.01 and an index recall of merely 0.11. These negligible scores underscore the inherent difficulty of the task. \revision{We additionally include a stronger baseline that prompts the same Qwen-3-32B model directly, without the intermediate components of Figure~\ref{fig:design-framework}. Although attaining the highest entry-recall, this baseline reached lower user-facing accuracy of 0.03, while the localisation metrics are also lower.}

Transitioning to the agentic workflow yields an improvement in user-facing accuracy for Gemma and Qwen. In particular, Qwen-3-32B exhibits the most significant gain, with user-facing accuracy rising from 4\% to 28\%. This implies that, based on empirical ground truth behaviour, developers would receive approximately one correct documentation update for every four recommendations with Qwen-3-32B and the agentic pipeline. While this increase comes at the cost of lower recall (potentially missing valid updates), it is accompanied by a substantial specificity increase (from 87.6\% to 98.7\%). This noise-filtering capability is important for minimising developer disruption in real-world workflows.


\revision{Furthermore, the agentic workflow yields better index localisation performance for two of the three models, and comparable performance for the third.} \revision{Specifically, the Qwen model improves its MRR from 0.68 to 0.73, while L2 Hierarchical Recall, identified as one of the levels (L1 and L2) capturing the most significant structural meaning~\cite{gao2025adapting}, rises from 0.87 to 0.90.} \revision{For Llama-3.3-70B the two workflows are on par, with the static configuration marginally ahead on index recall, 0.68 against 0.66.} \revision{Although exact index recall remains below 0.70, the hierarchical recall metrics reveal a distinct advantage from the agentic version of the pipeline.} \revision{The higher recall at L1 and L2 relative to exact index recall indicates that the framework more often identifies the correct high-level section than the exact paragraph within it, which is of some use to a developer scanning for the right region. We state the resulting operating point plainly, since it bounds what the framework can be used for. At its best configuration the framework does not detect about half of the PRs that were historically accompanied by a README update, and under the observed prevalence roughly three of every four recommendations it surfaces have no corresponding historical edit.
Part of that residue is documentation debt that developers genuinely overlooked, which we quantify in Section~\ref{sec:rq3}, and part of it is error.  The
framework is therefore a conservative triage mechanism that concentrates attention on a small subset of PRs, and it is not reliable enough for keeping a README up to date. We report it as evidence that the task is tractable enough to be worth pursuing, and Sections~\ref{sec:component_analysis} and~\ref{sec:rq2b} characterise what
currently stands in the way.}

\revision{Although exact index recall remains below 0.70, the hierarchical recall metrics show the agentic version more frequently locates the region containing the update. The high L1 and L2 recall, reaching 0.97 and 0.90 respectively for Qwen, means that when the model misses the exact paragraph it still directs developers to the correct high-level section, which is coarser than the task requires.} Therefore, based on the current performance of the pipeline, we recommend that it be used in a ``human-in-the-loop'' configuration, where humans validate the recommended updates or make their own documentation updates with assistance from the pipeline.

\revision{\subsection{Small-Model Performance and Computational Efficiency}}
\label{sec_small_model_efficiency}

\revision{To examine whether the pipeline remains practical with smaller models, we repeated the full end-to-end evaluation using Qwen-3-8B and Qwen-3.5-9B. Table~\ref{tab_small_model_efficiency} reports their detection and localisation performance together with token consumption, wall-clock latency, and estimated API cost. The Qwen-3-32B configuration is included as a reference point. We focus on the Qwen family here to enable a more direct comparison across model scales.}

\revision{The models were evaluated using different hardware configurations. Qwen-3-32B was served on two NVIDIA A100 80\,GB GPUs, while Qwen-3-8B and Qwen-3.5-9B were each served on one NVIDIA A100 80\,GB GPU. All configurations processed one PR at a time without cross-PR batching. The reported wall-clock times therefore represent the end-to-end latency of processing a single PR under each deployment setting. Since the hardware allocations differ, these measurements should not be interpreted as controlled comparisons of model inference speed. Expected token consumption, latency, and API cost are weighted using the 1:99 ratio introduced earlier. The monetary estimates describe an API-based deployment scenario rather than the cost of the local GPU infrastructure used in our experiments, and they assume no prompt caching. They use the OpenRouter prices listed on 31 July 2026. The input and output prices per million tokens were \$0.080 and \$0.280 for Qwen-3-32B, \$0.117 and \$0.455 for Qwen-3-8B, and \$0.100 and \$0.150 for Qwen-3.5-9B.}

\begin{table*}[t]
\centering
\small
\setlength{\tabcolsep}{4pt}
\caption{A comparison of the end-to-end effectiveness and computation efficiency of the agentic workflow for different model scales within the Qwen model family.}
\label{tab_small_model_efficiency}
\revisioncolor
\begin{tabular}{lrrrrrcrrrr}
\toprule
\multirow{2}{*}{Model}
& \multirow{2}{*}{ER}
& \multirow{2}{*}{Spec.}
& \multirow{2}{*}{UFA}
& \multirow{2}{*}{IR}
& \multirow{2}{*}{MRR}
& HR
& Exp. tokens
& Mean time
& Exp. time
& Cost \\
&
&
&
&
&
& (L1, L2, L3, L4)
& (in/out)
& pos./neg. (s)
& (s/PR)
& (USD/1,000 PRs) \\
\midrule
Qwen-3-32B   & 0.50 & 0.99 & 0.28 & 0.68 & 0.73 & (0.97, 0.90, 0.83, 0.81)  & 3,056/25 & 12.46/1.23 & 1.34 & 0.25 \\
Qwen-3-8B    & 0.39 & 0.97 & 0.11 & 0.68 & 0.73 & (0.96, 0.86, 0.73, 0.71) & 6,035/64 & \phantom{0}5.07/1.20 & 1.24 & 0.74 \\
Qwen-3.5-9B  & 0.21 & 0.99 & 0.22 & 0.75 & 0.80 & (0.97, 0.91, 0.84, 0.80) & 3,730/47 & \phantom{0}6.22/0.81 & 0.86 & 0.38 \\
\bottomrule
\end{tabular}
\end{table*}

\revision{Across the evaluated configurations, the mean latency remained below 13 seconds for positive PRs and below 1.3 seconds for negative PRs, and the highest observed p95 latency was 28.41 seconds. Weighted by the 1:99 ratio, the expected time per PR was 1.34 seconds for Qwen-3-32B, 1.24 seconds for Qwen-3-8B and 0.86 seconds for Qwen-3.5-9B. The pipeline therefore processes an individual PR on the scale of seconds under every configuration we evaluated, and moving to a smaller model shortens that time but still stays within the same scale.}

\revision{Using a smaller model does not translate to reduction of token consumption or API cost. Qwen-3-8B consumes roughly twice the expected input tokens of Qwen-3-32B, 6,035 against 3,056, because more entries proceed past the initial classification stage and into the generation and verification components. Under the listed OpenRouter prices the estimated cost is approximately \$0.74 per 1,000 PRs for Qwen-3-8B and \$0.38 for Qwen-3.5-9B, compared with \$0.25 for Qwen-3-32B. This ordering reflects the number of calls each configuration issues together with current provider-specific prices, and it may change as providers revise their pricing.}

\revision{The cost of a smaller model is instead borne in effectiveness. Qwen-3-8B reaches a user-facing accuracy of 0.11 and Qwen-3.5-9B of 0.22, against 0.28 for Qwen-3-32B. Qwen-3.5-9B attains a specificity comparable to that of Qwen-3-32B, but its entry recall falls to 0.21, so it misses most of the PRs that do require a README update. The localisation columns are not comparable across the three configurations, because each is computed only on the recommendations that the configuration delivered.
Qwen-3-32B delivers 5,358 of these, Qwen-3-8B 4,181 and Qwen-3.5-9B 2,225, so the smaller models are scored on progressively smaller and more selective subsets of the same positive set. Read in that light, Qwen-3-8B matches Qwen-3-32B on index recall and MRR while delivering roughly a quarter fewer recommendations, and the higher figures of Qwen-3.5-9B describe the minority of cases that survive an entry recall of 0.21. Overall, moving to a smaller model in this experiment lowers detection performance and degrades output well-formedness, while leaving localisation quality on the delivered subset unchanged, and in a commercial API setting it does not necessarily reduce price.}

\revision{To maintain a controlled comparison, each configuration used
the same model across all pipeline components. We did not combine
models of different scales within a single configuration, as this
would introduce an additional model-allocation variable and make
it difficult to separate the effect of model choice from that of
the pipeline design.}

\begin{tcolorbox}[left=1pt, top=1pt, right=1pt, bottom=1pt] \textbf{Answer to RQ1:} \revision{The agentic version reaches a specificity of 98.7\% and a user-facing accuracy of 28.0\% (RQ1a).} \revision{While pinpointing the exact index remains challenging, hierarchical recalls especially L1 and L2 are markedly higher, which indicates that the pipeline identifies the correct high-level section more often than the exact paragraph within it (RQ1b).} \revision{About half of the PRs that were historically accompanied by a README update are never surfaced, and roughly three of every four surfaced recommendations have no corresponding historical edit. Despite these limitations, the framework concentrates review effort on a small subset of PRs.}
\end{tcolorbox}

\section{Component Analysis (RQ2a)}\label{sec:component_analysis}

In this section, we analyse how the pipeline's specific component designs (C1--C5) address the dual challenges of context selection and severe class imbalance. Our objective is to identify the optimal configuration for each processing stage and to empirically demonstrate the necessity of the intermediate reasoning modules. We present each component's role and empirical performance alongside design justification. These justifications synthesise our findings to explain the rationale for our final architectural decisions.

\subsection{Methodology} 

A PR contains diverse information sources, including the title, description, commit messages, and file patches. We hypothesise that distinct pipeline components require varying forms of contextual information from the PR to function effectively. To validate this hypothesis, we conducted an ablation study on context configurations, systematically experimenting with input combinations ranging from high-level descriptions to comprehensive file patches. This analysis aims to identify the optimal information density for each stage of the pipeline.

We employ a hybrid evaluation strategy that synthesises quantitative automated metrics with qualitative human annotation. The \textbf{automated metrics} include: 
\begin{itemize}
    \item Entry Recall and Entry Specificity for the classification modules (C1 \& C5);
    \item Index Recall, MRR, and Hierarchical Recall for the localisation module (C4);
\end{itemize}

Complementing these metrics, \textbf{manual annotation} is used to validate the semantic integrity of the model's intermediate reasoning steps (C2--C3), which are opaque to automated metrics. We conducted manual inspection of a sampled subset of outputs, focusing on three critical dimensions: (1) the accuracy of the ``sufficiency'' determination (C2), (2) the relevance of the retrieved file patches (C3), and (3) the logical validity of the generated justifications.

To mitigate model-specific bias, we constructed a validation dataset using a majority voting paradigm. Entries were assigned a classification of ``Sufficient'' or ``Insufficient'' only if a consensus was reached by at least two of the three evaluated models (Llama-3.3-70B, Qwen-3-32B, Gemma-3-27B). From this consensus pool, we employed stratified random sampling~\cite{baltes2022sampling} to select 100 entries from the ``Sufficient'' branch and 100 entries from the ``Insufficient'' branch. Two authors independently annotated these 200 entries. We achieved substantial inter-rater agreement, with Cohen's Kappa scores of 0.74 for sufficiency classification, 0.80 for retrieved file validity, and 0.90 for justification validity, indicating a high level of agreement~\cite{mchugh2012interrater}. All discrepancies were resolved through subsequent discussion. Note that in this section, we present results for the static components in isolation to establish a clear baseline before discussing the agentic interactions.

\begin{table}[!t]
    \caption{Classification (C1) performance across different configurations (Abbreviation for context: \textbf{D}: Description, \textbf{C}: Commit Messages, \textbf{F}: File Names, \textbf{P}: Raw Patch Content.) }
    \centering
    \begin{tabular}{l l r r}
    \toprule
    Model   & PR Context & Entry Recall & Entry Specificity  \\
    \midrule
    \multirow{5}{*}{Gemma-3-27b}   & D & 53.6\%  & 90.0\%     \\
     & D + C & 55.9\%  & 91.3\%  \\
    & D + F & 62.5\%  & 83.9\%   \\
    & D + C + F & 68.5\%  & 83.8\%  \\
    & D + C + F + P & 79.9\%  & 50.9\%  \\
    \midrule
    \multirow{5}{*}{Qwen3-32B} & D & 70.3\%  &  69.4\% \\
    & D + C & 68.6\% & 77.8\% \\
    & D + F & 71.7\%  & 67.6\%  \\
    & D + C + F  & 72.3\% & 76.8\% \\
    & D + C + F + P & 79.8\% & 62.2\%  \\
    \midrule
    \multirow{5}{*}{Llama-3.3-70b} & D &  53.3\% & 73.5\%  \\
    & D + C & 73.6\% & 52.3\%  \\
    & D + F &  69.5\% & 61.5\%  \\
    & D + C + F  &59.0\% & 71.5\%   \\
    & D + C + F + P & 53.3\% & 75.3\%  \\
     \bottomrule
    \end{tabular}

    \label{tab:task1-results}
\end{table}

\subsection{Relevance Classifier (C1)}

\subsubsection{Component Description} 

\begin{table*}[!ht]
    \centering
    \caption{Localisation performance comparison for the sufficient and insufficient branches.}
    \begin{tabular}{l l c c c }
    \toprule
    Model &  Sufficiency & Index Recall & MRR & Hierarchical Recall\\
    \midrule     
        \multirow{2}{*}{Gemma-3-27b} & Sufficient (6,874) &0.62 & 0.67 & (0.95, 0.82, 0.70, 0.71) \\
        & Insufficient (3,828) & 0.41 & 0.47 & (0.93, 0.60, 0.48, 0.41)  \\
        \midrule
       \multirow{2}{*}{Qwen3-32B} & Sufficient (6,874)& 0.62 & 0.67 & (0.96, 0.85, 0.76, 0.76) \\
       & Insufficient (3,828) & 0.41 & 0.46 & (0.90, 0.64, 0.51, 0.45) \\
       \midrule
        \multirow{2}{*}{Llama3.3-70B} & Sufficient (6,874) & 0.66 & 0.70 & (0.96, 0.86, 0.75, 0.75) \\
        & Insufficient (3,828) & 0.42 & 0.46 & (0.92, 0.68, 0.55, 0.47) \\
        \bottomrule
    \end{tabular}
    
    \label{tab:sufficient-insufficient-compare}
\end{table*}

This initial component functions as a binary classifier, determining whether the repository's README necessitates an update in response to the changes introduced by a PR. The classifier outputs a definitive ``Yes'' or ``No'' decision. If ``Yes'' is predicted, the PR will be forwarded to the next step in the pipeline for processing. If ``No'' is predicted, we will discard the entry.

We prompted the LLM to perform this classification task, experimenting with varying levels of contextual information about the PR. Specifically, we evaluated the following five configurations: (1) PR description, (2) PR description and the list of commit messages, (3) PR description and the list of file patch names, (4) PR description, the list of commit messages, and the list of file patch names, (5) all the prior mentioned information plus the raw patch contents. Across all experimental settings, the pre-merge version of the raw README file was provided as a constant input.

\subsubsection{Component Performance}


As shown in Table~\ref{tab:task1-results}, the impact of increasing PR context varies significantly across models. For Gemma-3-27b, we observe a clear trade-off: providing richer context (especially raw patches) substantially improves recall (from 53.6\% to 79.9\%) but causes a severe drop in specificity (from 90.0\% to 50.9\%). This suggests that for some models, the abundance of low-level details introduces noise, causing the model to hallucinate connections in otherwise irrelevant PRs. However, this trend is less consistent for Qwen and Llama. For instance, Llama-3.3-70b achieves its highest recall with the lighter D+C configuration (73.6\%) rather than the full context, indicating that "more information" does not always equate to better performance. Despite these variations, the addition of raw patches (Configuration 5) tends to negatively impact specificity for the smaller models (Gemma and Qwen). In a production environment with high class imbalance (1:99), such a drop in specificity would result in an overwhelming volume of false positives, shifting an excessive filtering burden to downstream components.

Therefore, we identified the combination of \textbf{PR description, commit messages, and changed file names} as the optimal balance to apply in the final pipeline. In this setting, Recall increased for two models (Qwen and Llama) while specificity remained stable or improved. To avoid omitting instances that required README updates, we prioritised recall, provided that specificity does not degrade excessively. Therefore, we select this configuration as the one to apply in our pipeline, relying on subsequent components to refine the final recommendations delivered to developers.

\subsubsection{Design Justification} 

In real-world scenarios, the vast majority of PRs do not require README updates. Therefore, employing a lightweight classifier upfront significantly reduces the computational cost of the pipeline. However, due to the extreme class imbalance (approx. 1:99), even a classifier with high specificity admits a substantial number of false positives. Consequently, we designed C1 to act as a high-recall coarse filter, delegating the strict validation to the Rationale Examiner (C5) after detailed reasoning has been performed.

\subsection{Context Reasoning and Localisation (C2, C3, C4)} 

\subsubsection{Component Descriptions} 

In the static pipeline configuration, C2 (Sufficiency Analyser) determines whether the initial PR metadata (Title, Description, Commits) provides adequate context for accurate localisation. If the information is deemed insufficient, C3 (Retrieval Unit) executes a single-pass retrieval operation to fetch the most relevant file patch, notably without a verification step. Finally, C4 (Localisation and Generation Engine) processes the aggregated context, which comprises the metadata and any retrieved patches, to identify the specific README sections requiring updates and to generate the corresponding justifications.

In the agentic configuration, C2 and C3 operate within an iterative feedback loop. C3 employs a sliding context window of size $k$ (default $k=3$) to manage information retrieval. Initially, the top-$k$ file patches are retrieved and submitted to C2 for sufficiency assessment. If C2 deems the current context insufficient, C3 shifts the window to fetch the next subset of candidates (e.g., ranks 2 to $k+1$) and re-submits the data. This process repeats until sufficiency is achieved or a maximum of $p$ iterations (set to $p=3$) is reached.

Regarding the retrieval logic, C3 ranks file patches using a hybrid similarity metric that balances semantic relevance to the PR intent with alignment to the existing README content. The selection criterion for a patch $p_j$ is formulated as: 
\begin{equation}
    \operatorname*{arg\,max}_{j \in J} \left( \text{sim}(\text{desc}, p_j) + \max_{i \in I} (\text{sim}(\text{RM}_i, p_j)) \right)
\end{equation}
\noindent Where $J$ denotes the set of file patches within the PR, and $I$ represents the set of paragraph indices within the raw README file. 

Similarity is computed using cosine similarity based on embeddings generated by a lightweight MiniLM model~\cite{wang2020minilm}. The intuition for this strategy is to identify the file patch that best captures the intent of the PR while simultaneously exhibiting strong semantic relevance to the existing README structure, thereby providing the optimal context to facilitate accurate documentation updates.

\begin{table*}[!t]
    \centering
    \caption{``Sufficient'' entries localisation performance, comparing different levels of context.}
    \begin{tabular}{l l c c c }
    \toprule
    Model &  Context & Index Recall & MRR & Hierarchical Recall\\
    \midrule
       
        \multirow{3}{*}{Gemma-3-27b} &  PR title, description, and commits & 0.62 & 0.67 & (0.95, 0.82, 0.70, 0.71) \\
        & All Information & 0.56 & 0.61 & (0.95, 0.81, 0.66, 0.61) \\
        & Agent & 0.62 & 0.68 & (0.96, 0.84, 0.72, 0.74) \\
        \midrule
       \multirow{3}{*}{Qwen3-32B} & PR title, description, and commits & 0.62 & 0.67 & (0.96, 0.85, 0.76, 0.76)\\
       & All Information & 0.62 & 0.67 & (0.96, 0.85, 0.76, 0.74) \\
       & Agent & 0.64 & 0.70 & (0.97, 0.89, 0.79, 0.79)\\
       \midrule
        \multirow{3}{*}{Llama3.3-70B} &  PR title, description, and commits & 0.66 & 0.70 & (0.96, 0.86, 0.75, 0.75) \\ 
        & All Information & 0.66 & 0.70 & (0.94, 0.87, 0.76, 0.73) \\
        & Agent & 0.66 & 0.71 & (0.96, 0.87, 0.76, 0.73) \\
        \bottomrule
    \end{tabular}
    
    \label{tab:sufficiency-results}
\end{table*}

\begin{table*}[!t]
    \centering
    \caption{``Insufficient'' entries localisation performance, comparing different levels of context.}
    \begin{tabular}{l l c c c }
    \toprule
    Model &  Context & Index Recall & MRR & Hierarchical Recall\\
    \midrule
       
        \multirow{6}{*}{Gemma-3-27b} & None  & 0.41 & 0.47 & (0.93, 0.60, 0.48, 0.41)  \\
        & All Context Without Selection & 0.41 & 0.47 & (0.93, 0.66, 0.43, 0.41)\\
        & Top-1 File Patch & 0.44 & 0.50 & (0.94, 0.69, 0.55, 0.53) \\
        & Top-2 File Patch & 0.44 & 0.50 & (0.94, 0.70, 0.53, 0.51) \\
        & Top-3 File Patch & 0.41 & 0.47 & (0.93, 0.67, 0.52, 0.48)\\
        & Agent & 0.43 & 0.49 & (0.93, 0.69, 0.55, 0.52) \\
        \midrule
       \multirow{6}{*}{Qwen3-32B} & None &  0.41 & 0.46 & (0.89, 0.64, 0.51, 0.45) \\
       & All Context Without Selection & 0.43 & 0.48 & (0.90, 0.69, 0.58, 0.56)\\
       & Top-1 File Patch & 0.44 & 0.50 & (0.92, 0.70, 0.58, 0.53)\\
       & Top-2 File Patch & 0.43 & 0.48 & (0.92, 0.70, 0.57, 0.51)\\
       & Top-3 File Patch & 0.42 & 0.48 & (0.92, 0.70, 0.57, 0.49) \\
       & Agent & 0.47 & 0.52 & (0.95, 0.78, 0.64, 0.59) \\
       \midrule
        \multirow{6}{*}{Llama3.3-70B} & None  & 0.42 & 0.46 & (0.92, 0.68, 0.55, 0.47)\\
        & All Context Without Selection & 0.43 & 0.47 & (0.95, 0.72, 0.57, 0.50) \\
        & Top-1 File Patch & 0.45 & 0.50 & (0.94, 0.74, 0.61, 0.53) \\
        & Top-2 File Patch & 0.44 & 0.48 & (0.93, 0.72, 0.60, 0.49) \\
        & Top-3 File Patch & 0.42 & 0.46 & (0.93, 0.71, 0.55, 0.41)\\
        & Agent & 0.46 & 0.51 & (0.95, 0.76, 0.70, 0.45) \\
        \bottomrule
    \end{tabular}
    
    \label{tab:insufficient-results}
\end{table*}

\subsubsection{Sufficient Branch Performance} 

Upon determining that a PR requires a README update, the system first evaluates whether the existing metadata provides adequate context for identifying the specific sections requiring modification. If the information is deemed ``sufficient'', the component predicts the target locations and provides justifications to the developers explaining why the update is necessary. Otherwise, it directs the workflow to the insufficiency branch for further processing.

To evaluate this component in isolation, we utilised all data from the positive samples. We first tasked the LLMs with assessing the ``sufficiency'' of each entry. Out of 10,702 entries, Gemma, Qwen, and Llama classified 6,366, 6,752, and 7,341 entries respectively as ``sufficient''. The Krippendorff's alpha~\cite{krippendorff2018content} across the three models was 0.61, indicating a substantial level of agreement and suggesting that this task is suitable for LLM-based determination~\cite{ahmed2025can}. Furthermore, we conducted a manual validation on a random sample of 100 entries where all three models achieved consensus. We verified whether the PR metadata (title, description, and commit messages) provided adequate context to identify README updates needs. On this validated subset, the classification accuracy reached 83\%.

We then prompted the LLM to generate a list of indices for the sections requiring updates, along with justifications for each recommendation. Table~\ref{tab:sufficient-insufficient-compare} presents the localisation performance of three LLMs, comparing ``sufficient'' and ``insufficient'' entries using the PR title, description, and commit messages as context. For sufficient entries, the results demonstrate robust index recall performance at the paragraph level. Furthermore, the hierarchical evaluation indicates that the models are highly effective at identifying the correct top-level sections. Although performance declines when drilling down to leaf nodes, the prediction still maintains high recall over the top-level nodes, indicating the ability to offer recommendations close to the correct semantic region.

In contrast, localisation performance consistently declines across all metrics for ``insufficient'' entries compared to their ``sufficient'' counterparts. This performance disparity validates the sufficiency assessment---the sharp drop in accuracy confirms that the models correctly identified instances where the available context was truly inadequate for the task.

To validate that entries classified as ``sufficient'' truly contain adequate information for localising documentation updates, we benchmarked these entries against a baseline that utilised the full PR context (augmenting the metadata with all modified file patches). Table~\ref{tab:sufficiency-results} presents the results. For Gemma, adding the entire context significantly degrades performance, while the performance of the other two models is almost identical when presenting the extra contexts. This demonstrates that indiscriminately including all available content does not guarantee better localisation---rather, the surplus of low-level details introduces noise, overwhelming the models and hindering their ability to pinpoint the correct update targets.

Additionally, to ensure efficient inference on our hardware (2 A100 80G GPUs), we enforced a 32k token limit, truncating the tail of any inputs exceeding this threshold. This affected at most 3\% of the data with the longest context.  Excessive context length not only impedes inference speed, as attention complexity scales quadratically with sequence length~\cite{vaswani2017attention}, but also degrades instruction adherence. Empirically, we observed that providing uncurated context caused a breakdown in format compliance, with models failing to adhere the required generation schema in 17.3\% to 27\% of the ``sufficient'' entries. Therefore, providing the exhaustive context offered no additional benefit for accurately predicting update locations.

Finally, we manually inspected the justifications for the same subset of 100 randomly sampled entries. Validated under strong inter-rater reliability, the accuracy of the generated justifications reached 84\%. This high accuracy demonstrates the model's capacity to produce semantically sound reasoning, a critical factor for ensuring developer trust and facilitating the adoption of automated recommendations.

\subsubsection{Insufficient Branch Performance} 

For PRs lacking sufficient detail, we employ an auxiliary retrieval mechanism to augment available context. Leveraging the earlier content-based ranking strategy, we extract the file patch best capturing the PR intent while aligning with the existing README structure. This enriched context is then supplied to the model to facilitate accurate localisation.

Complementing our analysis of the sufficient branch, we performed a parallel manual verification for entries classified as ``insufficient''. This assessment confirmed an accuracy of 85\%, demonstrating that the models reliably detect when the available context is inadequate for localisation.

Table~\ref{tab:insufficient-results} details the experimental results for the localisation task on insufficient entries. We benchmarked the impact of retrieval by comparing a baseline of no additional context against strategies retrieving the top-$k$ file patches using the hybrid ranking metric. The results demonstrate that retrieving the single best-matching file patch (Top-1) provides the optimal information gain, with all metrics surpassing those of the Top-2 and Top-3 configurations. Furthermore, the exhaustive inclusion of all PR context fails to yield ideal performance compared to selective retrieval, underscoring the value of focused context augmentation.

Furthermore, we assessed the intermediate reliability of the retrieval and justification steps through manual annotation. We randomly sampled 100 entries classified as ``insufficient'' by at least two models. On this subset, Top-1 retrieval accuracy reached only 51\%, with justification accuracy at 55\%. These findings highlight the limitations of a static, single-round retrieval strategy. Often, the necessary information resides in a lower-ranked file (one with a lower match score) or requires the synthesis of a set of complementary file patches to fully elucidate the PR's intention. Crucially, while aggregate metrics suggest that increasing the number of retrieved files ($k$) degrades performance, this average obscures the heterogeneity of individual entries---specific PRs require distinct, dynamic retrieval configurations to resolve their ambiguity effectively.

This motivated us to design an agentic workflow capable of determining the specific information needs for each entry, and to conduct a further qualitative investigation into the failure modes distinguishing sufficient from insufficient cases. As shown in Table~\ref{tab:insufficient-results}, the agentic workflow improves index localisation for Qwen and Llama, while remains largely unchanged for Gemma. This improvement in two of the three models suggests that adaptively selecting the correct information in the appropriate quantity enhances localisation accuracy.

\subsubsection{Design Justification}

Our component-wise analysis informed two design decisions for the pipeline architecture:

\smallskip\noindent\textbf{Efficiency through Dynamic Context Selection:} A single PR often contains a vast amount of information. Our ablation study confirms that indiscriminately feeding this entire context (i.e., all file patches) significantly expands the input window, which not only increases computational cost but also degrades performance due to instruction-following failures (as seen in the ``Sufficient'' branch results). Conversely, the extreme class imbalance of README updates (approx. 1:99) means that deep processing for every PR is wasteful. Therefore, we designed the pipeline to be adaptive---C1 acts as a lightweight coarse filter to discard the majority of irrelevant PRs immediately, while C2 determines the minimum necessary context for the remaining PRs. This ensures that ``easy'' PRs are processed efficiently using only metadata, avoiding the noise and cost of processing raw code patches unless absolutely necessary.

\smallskip\noindent\textbf{Necessity of Iterative Retrieval:} Our evaluation of the ``Insufficient'' branch reveals that a single-pass static retrieval is often unreliable, achieving only 51\% relevance accuracy. Simply adding more static context (e.g., full patches) fails to improve localisation due to the ``lost in the middle'' phenomenon. These limitations explicitly justify the transition to an agentic workflow. Since static retrieval cannot reliably fetch the correct information in one shot, the system requires an iterative, self-correcting loop to dynamically navigate the file changes and locate the precise context needed for accurate updates.

\subsection{Quality Assurance Reviewer (C5)} 

\subsubsection{Component Description}

The final component in the pipeline functions as a validation gate, assessing the proposed update sections and their corresponding justifications against the current state of the README. This component processes the predicted update targets, intermediate justifications, raw README content, and full PR context, enforcing quality through two primary evaluation criteria:
\begin{enumerate}
    \item It verifies whether the proposed updates and justifications align with the existing README structure and the PR's intent, ensuring a coherent logical connection.
    \item It assesses whether the update's granularity is appropriate for the target section and confirms its necessity—determining if retaining the current version would render the documentation factually incorrect or obsolete. 
\end{enumerate}
Only recommendations that satisfy these conditions are presented to the developers.

\subsubsection{Component Performance} 

We manually investigated the justifications provided for false positives and observed that the majority exhibit an uncertain tone with vague suggestions. A characteristic of a README file is that it is not a comprehensive document enumerating every technical detail. While it is possible to associate codebase change with the documentation, it is crucial to maintain the level of detail present in the original version. Consequently, information that does not match the existing level must be excluded. For instance, a high-level ``Architecture'' section describing core design patterns should not be updated to list a minor internal utility library, as this detail disrupts the intended conceptual scope of the section.

To implement this filtering stage, we present the LLM with the raw README file alongside the suggestions and justifications generated in the previous step. Table~\ref{tab:rationale-examiner} details the performance of this review process. As shown, Gemma and Qwen struggle to correctly identify negative samples, failing to filter out invalid updates effectively. In contrast, Llama achieves a superior trade-off. Although it yields the lowest accuracy on positive samples, it achieves the highest accuracy on negative samples and offers the best balance. 

\begin{table}[!t]
    \centering
    \caption{Accuracy of the Rationale Examiner}
    \begin{tabular}{l r r}
        \toprule
        \textbf{Model} & \textbf{Positive} & \textbf{Negative} \\
         \midrule
         Gemma-3-27b & 77.5\% & 39.0\% \\
         Qwen3-32B & 68.1\% & 45.9\% \\
         Llama-3.3-70b & 62.4\% &  80.4\% \\
         \bottomrule
    \end{tabular}
    \label{tab:rationale-examiner}
\end{table}

In the agentic pipeline design, we divided this component into two distinct steps, focusing first on evaluating justification granularity and its alignment with the README. We hypothesise that the performance drop observed when processing excessive information stems from an information bottleneck, which prevents the model from effectively reasoning about the relationships between multiple artifacts. Conversely, we posit that a lack of concreteness in the provided justification often arises from insufficient context in the retrieved patch files. Therefore, we task this component with classifying the justification as ``correct'', ``hallucinating'', or ``generic''. Based on this classification, the system dynamically adjusts the retrieval window by increasing or decreasing the number of files whenever the justification is deemed unsatisfactory.

Subsequently, the second step focuses on assessing the validity of the README file itself. This phase determines whether the current documentation remains factually accurate despite the introduced code changes and evaluates the proposed justification for the update. By doing so, it functions as a stabiliser to prevent unnecessary edits to the documentation.

\subsubsection{Design Justification} 

In real-world scenario, the distribution of samples is highly skewed, with positive cases constituting only approximately 1\% of all PRs. Consequently, although the first component achieves high classification performance, developers would still be overwhelmed by a high volume of false positive recommendations. This component primarily serves as a guardrail to mitigate this issue. \revision{C1 decides from the PR alone whether a change is worth examining, whereas C5 judges a concrete proposed edit against the README text, which does not exist before C4 and therefore cannot be assessed earlier. In the agentic Qwen-3-32B configuration, 7,524 positive PRs reach this stage and 28.8\% are discarded, and those discarded outputs reach an index recall of 0.51 against 0.68 for the ones delivered. On the negative set the check removes 383 of the 576 PRs that reach it, which leaves 193 false positives and gives the entry specificity of 0.99 reported in Table~\ref{tab:end-to-end}. Without it, entry recall would rise to 0.70 and specificity fall to 0.96, giving a user-facing accuracy of 0.15 rather than 0.28.}
Furthermore, within our agentic pipeline, we incorporated the capability to dynamically adjust the amount of extracted information based on self-reflection of the model's intermediate reasoning, thereby improving end-to-end performance.

\begin{tcolorbox}[left=1pt, top=1pt, right=1pt, bottom=1pt] 
\textbf{Answer to RQ2a:} Our pipeline addresses context selection by decomposing the task into staged components that progressively filter, enrich, and validate information, and it addresses severe class imbalance by combining a high-recall upfront filter with a strict downstream quality assurance layer to suppress false positives. Our analysis confirms the superiority of selective context over indiscriminate inclusion. Metadata alone proves optimal for well-described PRs, avoiding the noise of raw patches, whereas complex cases necessitate the dynamic, iterative retrieval of an agentic workflow. Furthermore, the extreme class imbalance demands a dedicated quality assurance layer to filter false positives and ensure documentation stability.
\end{tcolorbox}

\section{Failure Cases Analysis (RQ2b)}
\label{sec:rq2b}

Complementing the automated component evaluation, we conducted a qualitative analysis of failure cases where our pipeline failed to localise the ground truth update targets. This examination aims to characterise the inherent complexity of the task and delineate the limitations of our current approach, thereby identifying critical directions for future improvement.

\subsection{Methodology} 

We drew samples from the same ``sufficient'' and ``insufficient'' data pools established for the intermediate evaluation. To specifically examine failure patterns, we extended our random sampling independently within each branch to isolate 200 distinct instances (100 from the ``Sufficient'' pool and 100 from the ``Insufficient'' pool) where the models failed to capture the complete set of ground truth indices. This selection explicitly excluded cases of perfect localisation. 

For these samples, we generated outputs using all three LLMs. To isolate failure cases from model variability or accidental errors, we applied a majority voting mechanism---an index was considered ``predicted'' only if identified by at least two models. Consequently, we defined overlooked indices as ground truth sections that failed to achieve this majority consensus, ensuring our analysis focuses on persistent, model-agnostic errors over random noise. The qualitative analysis was then conducted based on the overlooked indices. 
 
We subsequently performed a root cause analysis on the indices overlooked by the models. To ensure a systematic derivation of failure causes, we adopted a hierarchical investigation protocol with a bounded time window, adapting the approach of Gao et al.~\cite{gao2025adapting}. This protocol was iteratively refined based on pilot data to systematically isolate procedural and informational deficiencies while imposing strict boundaries on investigation depth, acknowledging that the authors, as external observers, possess finite project-specific knowledge. The procedure entailed the following sequential steps: 
\begin{enumerate}
    \item Inspecting the specific missed README section;
    \item Analysing the PR title, description and commit messages to infer the PR intent;
    \item Examining the retrieved files (exclusively for the ``Insufficient'' branch);
    \item Determining whether the target content relied on external context absent from the examined fields;
    \item Evaluating the logical alignment between the documentation update and the code change intent;
    \item Assessing the README structure to determine if the update was inferable by a human observer; 
    \item Checking for project-specific knowledge that could hinder generalisation.
\end{enumerate}
We processed each index sequentially, enforcing a five-minute limit per step. The investigation terminated immediately upon identifying a distinct cause. Conversely, if all steps were exhausted or the time limit expired without a clear finding, the case was classified as having no obvious human-identifiable reason. For each entry, this process yielded a concise textual description characterising the specific barrier to localisation.

To synthesise these unstructured failure descriptions into a coherent taxonomy, we first employed open coding~\cite{glaser2017discovery} to derive granular base-level codes. The initial coding schema was established using a set of 20 samples, which were independently analysed by two researchers to resolve discrepancies and calibrate the criteria. Following this phase, the lead author processed the remaining samples, flagging ambiguous entries for adjudication during weekly team meetings. Subsequently, we applied open card sorting~\cite{zimmermann2016card} to structure these codes. This involved an iterative process of clustering the base-level codes according to semantic similarity to identify high-level emerging themes. This multi-stage collaborative review served as a rigorous quality control mechanism, ensuring the consistency and validity of the final failure categories.

\subsection{Results} 

We identified 12 types of failure cases, which are further merged into 4 higher-level categories, namely (1) context and information deficit, (2) semantically irrelevant noise, (3) document structural ambiguity, and (4), all other failures.

\subsubsection{Context and Information Deficit}

This category encompasses cases where the input information provided to the models is insufficient for deriving the correct update location. It accounts for 34\% and 32\% of the analysed failures in the sufficient and insufficient branches, respectively.

\smallskip\noindent\textbf{Vague or insufficient PR description (15 occurrences in the ``sufficient'' branch, 9 occurrences in the ``insufficient'' branch):} This issue is more prevalent in the ``sufficient'' branch, where oversimplified descriptions obscure the PR's intention. Consequently, without external context, the PR information alone is inadequate for deriving the target indices. For the insufficient branch, the PR description remains indispensable: it is required to provide a high-level summary of the intention that the retrieved file patches alone cannot convey. For example, one assessed PR~\cite{example1} uses informal language, stating ``the split was starting to really annoy me'', without clarifying the relevant terminology.

\smallskip\noindent\textbf{Ineffective File Retrieval (0, 8):} This issue is exclusive to the ``insufficient'' branch. It arises when a vague PR description is compounded by the retrieval of an irrelevant file patch. In the absence of useful context from either source, the correct indices cannot be inferred.

\smallskip\noindent\textbf{Unreachable External Resources (7, 10):} This issue pertains to inaccessible content such as figures or linked PRs. Since the model cannot parse visual data within figures or retrieve information from other PRs, it fails to make decisions based on that context. For example, a sampled PR~\cite{example2} updates the contributor list; however, the model overlooks that the ``contributor count'' icon also requires an update, as it cannot read the number embedded in the figure. Most failures in this category could be mitigated by enabling access to these external resources.

\smallskip\noindent\textbf{Project-Specific Knowledge Gap (14, 12):} Failures in this category stem from a lack of deep project-specific knowledge, such as historical development patterns, the overall project structure, auxiliary files, or metadata. For example, a sampled PR~\cite{example3} indicates a need to update the installation section; however, the impact is strictly limited to the Ubuntu system configuration. Without access to broader context, such as related files or the specific directory structure, the model fails to identify this constraint.

\subsubsection{Semantically Irrelevant Noise}

This category attributes failures to ``noise'' stemming from inconsistent or arbitrary developer patterns rather than model incapacity. These failures are inherently difficult to predict and suggest a need for more atomic PRs and stricter commit discipline. This category accounts for 40\% and 44\% of the failures in the sufficient and insufficient branches, respectively.

\smallskip\noindent\textbf{Document Cosmetic Changes (23, 22):} Failures in this category involve minor editorial adjustments, such as correcting typos, rephrasing sentences, or fixing Markdown syntax. Since our study primarily focuses on documentation updates triggered by code changes, these cosmetic edits fall outside our scope. Furthermore, this is a relatively trivial challenge; specialised formatting tools, such as the template-based approach proposed by Gao et al.~\cite{gao2025adapting}, can effectively resolve these issues.

\smallskip\noindent\textbf{Fixes to Past Document Issues (2, 7):} Failures in this category involve retroactive corrections to pre-existing documentation issues. While the code changes in the current PR represent new development, the documentation updates respond to past code changes that lie outside the scope of the current PR. Consequently, these updates act as noise, as the current PR provides no evidence to predict them. This discrepancy highlights a latency in synchronising documentation with rapid code evolution, which our proposed pipeline aims to address. For example, a sampled PR~\cite{example5} explicitly addresses a fix for a separate PR merged 14 days prior.

\smallskip\noindent\textbf{Tangled Pull Requests (0, 7):} This category encompasses PRs that bundle multiple distinct development activities, effectively serving as an aggregation of unrelated tasks. It also includes routine procedural documentation maintenance. In such cases, the model is inundated by the massive and diverse information load. Furthermore, this complexity is often compounded by oversimplified PR descriptions, which leads to unreliable retrieval performance and prevents the model from accurately predicting the necessary updates.

\smallskip\noindent\textbf{Erroneous Ground Truth (0, 3):} Failures in this category arise from inaccuracies within the ground truth, where the documentation updates introduced by the developers contained errors or inconsistencies. Since the model attempts to predict a logical update based on the PR context, it cannot reproduce these human errors, resulting in an inevitable mismatch.

\smallskip\noindent\textbf{Other Misalignment Between PR Intent and Document Update (18, 14):} These failures occur when document updates deviate from the PR's primary purpose. These indices are unpredictable due to a lack of supporting evidence within the PR context; instead, they appear to stem from arbitrary or ad-hoc developer behaviour. For example, a sampled PR~\cite{example4} focused on improving code-level logging, yet the documentation was updated to include a new group chat channel, a change clearly divergent from the PR's overarching goal.

\subsubsection{Document Structural Ambiguity}

This category encompasses failures where the specific structure of the README file impedes accurate index prediction. This category accounts for 23\% and 24\% of the failures in the sufficient and insufficient branches, respectively.

\smallskip\noindent\textbf{Ambiguity in Inserted Location (17, 18):} Failures in this category stem from the existence of multiple plausible locations for inserting new content. While the model's reasoning demonstrates a correct understanding of the PR's intention, it occasionally suggests an insertion point different from the ground truth. These discrepancies do not necessarily indicate an incorrect answer, as inserting the content at these alternative locations often preserves the document's coherence. For example, in a sampled PR~\cite{example6}, the model suggested a location for a new API client section. Given that the README contained a list of similar sections, placement at any of these positions would remain contextually valid.

\smallskip\noindent\textbf{Coarse-Grained Localisation (7, 9):} Failures in this category occur when the model identifies the correct section but conservatively predicts the update at the section header rather than the specific content location within that section. Despite the coarse granularity, the model's justification confirms a correct understanding of the PR. This limitation likely stems from our preprocessing strategy, which represents the README as a flattened list of sections, causing the model to occasionally lose track of internal substructures. To address this, we employed a finer-grained hierarchical evaluation to account for these semantically accurate but structurally coarse predictions.

\subsubsection{Other Failures}

This final category (12, 6) encompasses failures where the root cause could not be identified despite adhering to the systematic analysis protocol outlined in the methodology, or where the manual diagnosis exceeded the pre-determined time limit.

\begin{tcolorbox}[left=1pt, top=1pt, right=1pt, bottom=1pt] 
\textbf{Answer to RQ2b:} Over 40\% of failures are driven by Developer-Induced Noise (e.g., cosmetic fixes, tangled PRs, and ad-hoc updates) rather than model reasoning deficits. This indicates a theoretical ceiling for automation in ``wild'' PRs: without stricter contribution guidelines, many updates are simply unpredictable. Meanwhile, roughly 24\% of the failures are related to ambiguity in the document structure. However, roughly 33\% of failures (Context Deficit) are addressable by incorporating corresponding information, enabling pipeline improvements in future work. 
\end{tcolorbox}

\section{Detecting Overlooked Updates (RQ3)}
\label{sec:rq3}

As established in our previous experiments, the inherent class imbalance, characterised by a high volume of negative samples, causes our tool to generate recommendations for PRs that lack a corresponding ground-truth README update. In this section, we investigate whether a subset of these ostensibly ``false'' positives actually reflect necessary documentation updates overlooked by the developers---as also observed in related research on LLM-based test maintenance~\cite{liu2025llmtestmaintenance}. Furthermore, we evaluate how developers perceive and manage the genuine false positives generated by our tool.

\subsection{Methodology} 

We conducted a historical analysis on 20 repositories: 19 randomly sampled from our dataset and one additional repository \texttt{JabRef/jabref}~\cite{jabref} where we established contact with a maintainer. We executed our pipeline across the entire PR history of these projects, which averaged 8,379 PRs per repository, to identify PRs that potentially required a README update but lacked one at the time of closure.

Validating the necessity of every README edit recommendation would require deep, project-specific domain knowledge that external researchers lack, rendering a comprehensive manual analysis both time-consuming and prone to subjective bias. To mitigate this subjectivity and focus on high-confidence candidates, we employed a proxy validation strategy based on subsequent development activity. Specifically, we applied a time-window validation, scanning the README commit history for one month following the closure of each flagged PR. If the README was subsequently updated in the exact section predicted by our tool during this window, we classified the PR as a potential ``delayed fix.''

To validate that these delayed updates truly corresponded to the PR's intent rather than coincidental edits, we conducted a manual inspection. We randomly sampled up to 10 ``delayed fix'' candidates per repository for the 19 sampled repos, and annotated all 18 candidate PRs for the \texttt{jabref} repository. In total, we manually reviewed 163 entries, classifying them into three categories:
\begin{itemize}
    \item \textbf{Delayed Fixes:} The recommendation matched a later README update with the same intent;
    \item \textbf{Valid-but-Ignored:} The recommendation was reasonable and necessary, but the README remains outdated;
    \item \textbf{Coincidental/Irrelevant:} The later update was unrelated to the PR.
\end{itemize}

Finally, for the \texttt{jabref} repository, we sampled four PRs containing potential delayed fixes (three confirmed as delayed fixes and one classified as valid-but-ignored), along with four additional cases lacking subsequent matches. \revision{This part of the study is a formative assessment with one maintainer of one project on eight recommendations, half of which had already been confirmed by a temporal match, and it was neither blinded nor randomly sampled.
It is therefore reported as anecdotal evidence of how the recommendations are received, not as a measure of how often they are correct.} In a one-hour in-person interview, we evaluated the validity of the recommended updates and the soundness of the accompanying justifications, followed by a series of open-ended questions, including 1) the acceptable interaction pattern with the tool, 2) the use case scenarios of the tool, 3) the sentiment towards false positive samples, and 4) the applicability and implications of the tool in other types of document artefacts.

\subsection{Results} 

\begin{table}[t]
\revisioncolor
\centering
\caption{Funnel of the retrospective analysis.}
\label{tab:rq3-funnel}
\begin{tabular}{l r}
\toprule
Stage & Count \\
\midrule
Repositories analysed & 20 \\
Pull requests in their full history & 167,580 \\
Flagged, README not updated at closure & 1,832 \\
Predicted section updated within one month & 425 \\
Manually annotated & 163 \\
\midrule
\quad Delayed fix & 25 \\
\quad Valid but ignored & 10 \\
\bottomrule
\end{tabular}
\end{table}

\revision{Table~\ref{tab:rq3-funnel} summarises the overall funnel.} Across the 20 repositories, our tool flagged 1,832 PRs, averaging \revision{91.6} per repository, as requiring updates that were absent at the time of closure. Applying the one-month latency filter reduced this count to 425 PRs, averaging 21.3 per repository, where the specific section was later updated.

Our manual annotation of 163 temporal matches revealed the tool's precision in identifying genuine documentation debt. Specifically, we confirmed that, in 25 cases representing approximately \revision{15.34\%} of the sample, the tool correctly identified a necessary update that was only addressed by developers in a subsequent, separate commit. Furthermore, we identified 10 instances where the tool's recommendation was technically accurate and necessary, even though the documentation remains outdated. These cases represent ``silent'' technical debt successfully surfaced by our pipeline. \revision{The temporal heuristic captures some noise, and the two categories differ in their reliance upon it. The 25 delayed fixes assume the earlier PR prompted a later README edit, whereas the 10 unaddressed cases lack later fixes. We therefore treat the 10 as a conservative figure and the 35 together as an upper bound.}


In the targeted evaluation of the \texttt{jabref} repository, the maintainer confirmed that six out of eight recommendations were valid and warranted documentation updates. Notably, this subset included all four cases identified through temporal matches, which we had previously classified as three ``delayed fixes'' and one ``valid-but-ignored'' update. Furthermore, the maintainer acknowledged that two of the four cases lacking any subsequent historical matches were genuine oversights, where the documentation should have been updated according to our tool's recommendation.

Regarding the two negative cases where the maintainer concluded that no update was necessary, one recommendation proposed updating the licence section regarding a legal constraint on new functionality, while the other suggested a modification to a function description. For the first case, the maintainer reflected: \textit{``It is tough as it is a legal issue, and related to risk management. So the team need to decide whether to take the risk... However, it (the recommendation) is a good tip, but it does not change the general licence per se, and the change is actually in the external libraries, so we do not want to change it''. } 

Regarding the second case, while the update was deemed unnecessary, the developer noted the low overhead and supportive nature of the tool: \textit{``It took me 20 seconds, and back then it would only take me 15 seconds. This is quite lightweight because you indicated the location and some explanation and this would be OK. This is actually supportive as it increases documentation quality directly or indirectly. By indirectly I mean maybe this will raise an issue in the future [or] reminds the part of the README that should be updated''}. 

Therefore, even recommendations deemed unnecessary for an immediate update serve a valuable purpose: they either pinpoint locations requiring deeper team discussion or act as non-intrusive reminders that assist developers in reviewing corresponding documentation sections.

\begin{tcolorbox}[left=1pt, top=1pt, right=1pt, bottom=1pt] \textbf{Answer to RQ3:}  Our analysis of 163 cases with temporal matches confirmed 21.5\% as valid recommendations overlooked by developers (comprising 15.34\% delayed fixes and 6.13\% valid but ignored). \revision{Feedback from a single jabref maintainer is consistent with this, with six out of eight recommendations judged to be genuine oversights, two of them without temporal matches.} 
\revision{These findings indicate that some recommendations without a matching historical edit correspond to real documentation debt rather than to errors, which places the user-facing accuracy reported in Table~\ref{tab:end-to-end} at the low end of what a deployment would see.}
\end{tcolorbox}

\section{Discussion and Implications}

\subsection{Implications for Practitioners}

This study offers concrete implications for practitioners. Our results highlight the prevalence of outdated documentation: while only approximately 1\% of PR activity involves document updates, our retrospective analysis demonstrates that our tool successfully detected necessary updates that were either ignored (and later remedied) or entirely overlooked (as confirmed by an OSS maintainer). This finding echoes previous evidence~\cite{aghajani2019software} regarding the severity of documentation up-to-dateness issue. Consequently, practitioners should prioritise documentation verification during code reviews.

In a purely manual workflow, developers must screen every PR to identify the rare instances that require documentation updates. Given the sparsity of such cases, this constant vigilance creates a high-effort, low-reward dynamic. By incorporating our approach into the PR workflow, this reviewing effort is significantly concentrated. As indicated in Table~\ref{tab:end-to-end}, the best-performing approach achieves a user-facing accuracy around 28\%. This metric provides a crucial perspective on the tool's practical utility relative to the baseline human effort: the model ensures that for approximately every three to four automated recommendations reviewed, the developer will identify one genuine necessity for a README update. This ratio transforms documentation maintenance from an intractable manual burden into a manageable operational task.

Our tool fits naturally into CI/CD pipelines, functioning as a documentation bot to monitor the README file. As noted by the \texttt{jabref} developer, the fine-grained, paragraph-level recommendations, accompanied by justifications, draw necessary attention to documentation needs with minimal distraction to the primary PR workflow. The developer further suggested that generating concrete edit candidates at pinpointed locations would be a valuable future enhancement. However, they emphasised that such suggestions must be verified and finalised by humans. The developer contrasted this human-in-the-loop approach with the alternative of regenerating the entire README from scratch for every PR, stating: \textit{``There are a range of README files ... there should always be human involved to do prioritisation, and to finally polish the README. Because we are not talking about toy projects, we are talking about projects wanting to have user community, and we want to talk to the users who are human, so I don't believe this can be entirely generated by AI.''}

This underscores the importance of reserving the final editorial decision for human developers to ensure that documentation remains accurate and consistent with a project's specific technical voice. With the proliferation of LLMs, such tools are becoming increasingly ubiquitous in Open Source Software (OSS) projects, gradually assuming the role of ``AI teammates''~\cite{li2025rise}. Consequently, for projects addressing documentation debt via delayed updates, future documentation bots targeting multiple artefacts could automatically flag stale content as issues, allowing developers to batch these tasks and delegate initial drafting to AI agents. Developers would then perform fine-grained, human-in-the-loop adjustments to verify the content. This workflow exemplifies the emerging paradigm of collaborative human-AI software engineering.

\subsection{Implications for Researchers}

For researchers, our proposed workflow serves as a proof-of-concept, demonstrating that automated, reasoning-based interventions are a viable solution for maintaining documentation up-to-date. However, the current approach faces challenges regarding the imbalanced distribution of PRs requiring README updates, alongside information bottlenecks.

For failure cases in localising update indices, our failure analysis identifies ``semantically irrelevant noise'' as the primary source of unpredicted cases, while ``document structural ambiguity'' ranks third. Regarding the former, ground-truth updates are often arbitrary or purely cosmetic; this renders them unreliable as a gold standard for automated evaluation. Regarding the latter, ground-truth indices often reflect individual developer preferences, despite the frequent existence of multiple equally valid insertion points. Notably, our best-performing approach aggressively filters out nearly half of the PRs originally labelled as ``positive'' in the dataset. However, our qualitative analysis reveals that when we force the model to generate indices for the entire set of ``positive'' PRs, a significant proportion of these failures is rooted in semantically irrelevant noise or arbitrary developer preferences. Consequently, the practical recall rate is likely higher than what is reflected by automated metrics, as the model correctly ignores PRs where the recorded human update was unnecessary or cosmetic.


\revision{Regarding the PRs that developers did not update back then, our investigation into flagged ``negative'' PRs, where the tool recommended an update contrary to the ground truth, indicates that a measurable share of them were not errors.} Among cases identified via temporal matches, 21.5\% were confirmed as necessary updates overlooked by developers, representing actual instances of documentation drift. This finding is further corroborated by our \texttt{jabref} case study; even among randomly sampled recommendations lacking subsequent temporal matches, the maintainer confirmed that two out of four represented genuine oversights. \revision{A share of the ``false positives'' penalised by the automated evaluation therefore corresponded to documentation work that was outstanding, which bounds how much of the measured specificity gap is model error, echoing similar findings in research on LLM-based test maintenance prediction~\cite{liu2025llmtestmaintenance}.}


In the subsequent open discussion regarding desired features, the developer emphasised that the tool should function within the CI/CD pipeline and provide feedback directly as PR comments, noting: \textit{``If your output has line numbers and the reasons, then it is a no-brainer, and that is what I want to see.''} Regarding the tolerance for false positives, the developer expressed a pragmatic view: \textit{``I would expect two-thirds to three-quarters to be OK, and the rest being rubbish. But if the tool spots something that we did often... rubbish is also acceptable.''} Finally, addressing the grey area where a valid update might be withheld due to preference, the developer remarked: \textit{``This would not be distracting, as if I see my documentation bot is making a review comment, I can ignore it by simply closing it, because the paragraph locations are ranked, and I am not looking at kilobytes of text but a few short paragraphs.''}

Turning to the challenge of information bottlenecks, the second most common failure category is ``context and information deficit''. This stems largely from the constraints of a fixed evaluation pipeline, which requires the model to generate outputs based on a pre-determined and potentially insufficient set of retrieved information. Notably, performance improves under our agentic pipeline, where the system is empowered to reason over available data and prioritise context before finalising a decision. This confirms that LLMs can facilitate more informed documentation updates when given the agency to resolve information gaps autonomously. A comparison of the sufficient and insufficient branches indicates that failure volumes across most categories are largely similar. However, the underlying failure patterns reveal distinct vulnerabilities: the sufficient branch is more susceptible to inadequate PR descriptions, whereas the insufficient branch, despite its ability to retrieve supplementary files, remains prone to failures originating from the retrieval process itself. This divergence provides a compelling motivation for the model to self-reflect on its decisions.

Therefore, future work could further enhance this capability by incorporating external information and fine-grained project metadata through organised tool calls. By integrating document formatting and rephrasing capabilities, the system could mitigate identified cosmetic failures and reduce invalid updates, such as retroactive fixes to outdated documentation. Furthermore, employing improved documentation representations that preserve the hierarchical structure would also enhance the model's reasoning effectiveness. Finally, while this study focused on README files which serve the generic entry point for most repositories, the pipeline could be adapted to cover other documentation types, such as user documents, API documents, and contributing guidelines. 
\revision{This is not merely a matter of pointing the pipeline at another file, since relevant documentation may span multiple artefacts and may even be maintained outside the repository~\cite{milliken2025beyond}. Handling such artefacts gracefully would largely leave the pipeline components unchanged while generalising their input, so that a PR is first routed to the documents it could plausibly affect and the same localisation process is then applied to each of them, with periodically refreshed snapshots providing access to documentation hosted outside the repository.}
Future work should also explore alternative interaction paradigms on a larger scale to identify the optimal socio-technical workflows for maintaining documentation in the open-source ecosystem.

\revision{Our evaluation uses the same model across all pipeline components. A heterogeneous configuration could instead assign a smaller model to the initial binary classification task and reserve a larger model for subsequent retrieval, generation, and verification. Such an allocation may offer a different balance between effectiveness, latency, and cost, but it also introduces deployment trade-offs. In our current serving setup, Qwen-3-32B requires two A100 80\,GB GPUs and Qwen-3-8B uses a separate A100 80\,GB GPU. Keeping both services active would therefore increase the concurrent allocation from two to three GPUs. Sequentially loading the models could avoid this additional allocation but would add model-loading latency and operational complexity. Future work could also evaluate component-specific model routing as a joint optimisation of effectiveness, latency, token consumption, and hardware use.}

\section{Threats to Validity}

\noindent\textbf{Construct Validity: } The construct in this study implicitly requires that the document updates were triggered by the co-occurring codebase modifications, as we rely on the dataset of PRs with README updates in it.
\revision{We do not claim that this holds for every entry, and our failure analysis and retrospective analysis both show that it does not. What our qualitative and retrospective analyses establish is narrower, namely that the deviations are of identifiable kinds and that the direction of the resulting error is not uniform.}

\smallskip\noindent\textbf{Internal Validity: } First, our end-to-end evaluation relies on the ground truth of historical development activities. However, our failure analysis indicates that a substantial percentage of miss-predictions stem from inherent flaws in this ground truth, such as misalignment between documentation updates and the actual codebase changes, or structural ambiguities in the README that render multiple update locations plausible. Second, as demonstrated in our retrospective analysis, many recommendations classified as ``false positives'' were actually necessary updates that were either overlooked (and fixed later) or explicitly acknowledged by the developer. Therefore, our reported performance metrics likely represent a conservative lower bound of the tool's actual utility. Third, we conducted the evaluation on the complete positive dataset to facilitate a fair comparison between different models. In a real-world deployment, certain PRs might be pre-filtered based on heuristics, potentially causing performance variations. \revision{Finally, a README update observed within the one-month window is not by itself evidence that the originating PR overlooked anything, since some projects batch documentation work or defer it to release preparation. The conservative reading of our retrospective analysis therefore rests on the ten valid-but-ignored cases, or 6.1\%, with the 21.5\% that also counts delayed fixes as an upper bound.}

\smallskip\noindent\textbf{External Validity: } First, we evaluated our proposed approach across nine popular programming languages; consequently, we do not claim generalisability beyond this scope. Second, this study focused on README files, the high-level introductory documentation common to most OSS repositories. While this pipeline could extend to other documentation artefacts, results may vary, and the approach would require adaptation to suit different document structures. Regarding usability, we relied on an in-depth interview with a core developer from \texttt{jabref}. As this single perspective may not represent the entire OSS developer population, future work could conduct broader user studies or large-scale surveys to validate these findings across a diverse range of projects. Finally, exact performance reported in the paper is a result of specific LLMs and their versions. As models evolve, performance might change.

\section{Conclusion}

\revision{In this study, we formulated surgical documentation update recommendation as a task and presented a framework that instantiates it in a human-in-the-loop workflow.
Given a PR, the framework detects whether a README update is required, and if so, it pinpoints the paragraph-level locations and provides a justification linked to the triggering events within the PR to assist developers.}

Through both end-to-end and component-wise evaluations against historical development activity, we found that our best-performing model yields one valid recommendation for every four suggestions. Additionally, we conducted a qualitative analysis of failure cases where the tool failed to predict the ground truth indices. We identified that the primary failure scenarios stem from semantically irrelevant noise, contextual and informational deficits, and structural ambiguities within the documentation. Finally, we performed a retrospective analysis on instances where our approach recommended updates despite the absence of a corresponding ground truth change. In our analysis of recommendations that matched subsequent edits, manual annotation confirmed that 21.5\% identified necessary updates that were overlooked at the time. \revision{An in-person interview with one OSS developer offers anecdotal support, with recommendations judged valid in two out of four cases that lacked any later matches.}

\revision{Taken together, these findings establish the feasibility of surgical documentation update recommendation and delimit how far a current LLM-based approach reaches on it.} We concluded by outlining concrete implications for developers and researchers, specifically highlighting the potential for expanding this approach to other documentation artefacts and experimenting with diverse interaction paradigms within the OSS development workflow.

\section{Replication Package}
Our replication package is available at \url{https://github.com/Haoyu-Gao/README-auto-update}.

\appendix[Component used for computing the metrics]

\revision{We add the components of the entry-level classification metrics in Table~\ref{tab:confusion}}
\label{append-component}

\begin{table}[t]
\revisioncolor
\centering
\caption{Components of the entry-level classification metrics.}
\label{tab:confusion}
\begin{tabular}{l l | r r r r}
\toprule
Model & Version & TP & FP & TN & FN \\
\midrule
\multirow{2}{*}{Gemma-3-27b} & Static & 5,635 & 1,441 & 13,368 & 5,067 \\
 & Agentic & 7,421 & 921 & 13,888 & 3,281 \\
\midrule
\multirow{2}{*}{Qwen-3-32b} & Static & 5,231 & 1,840 & 12,969 & 5,471 \\
 & Agentic & 5,358 & 193 & 14,616 & 5,344 \\
\midrule
\multirow{2}{*}{Llama-3.3-70b} & Static & 4,073 & 817 & 13,992 & 6,629 \\
 & Agentic & 5,815 & 1,601 & 13,208 & 4,887 \\
\midrule
Qwen-3-32b & Baseline prompt & 8,198 & 4,383 & 10,426 & 2,504 \\
\midrule
Qwen-3-8b & Agentic & 4,181 & 492 & 14,317 & 6,521 \\
Qwen-3.5-9b & Agentic & 2,225 & 109 & 14,700 & 8,477 \\
\bottomrule
\end{tabular}
\end{table}

\bibliographystyle{IEEEtran}
\bibliography{reference}

\end{document}